# Bayesian Origin-Destination Estimation in Networked Transit Systems using Nodal In- and Outflow Counts

Steffen O.P. Blume[a,b], Francesco Corman[c], Giovanni Sansavini[a]

[a]Reliability and Risk Engineering Laboratory, ETH Zürich, Leonhardstrasse 21, 8092 Zürich, Switzerland; [b]Future Resilient Systems, Singapore-ETH Centre, 1 Create Way, CREATE Tower #06-01, Singapore 138602; [c]Institute for Transport Planning and Systems, ETH Zürich, Stefano-Franscini-Platz 5, 8093 Zürich, Switzerland

Corresponding author: Giovanni Sansavini, sansavig@ethz.ch

**Abstract**
We propose a Bayesian inference approach for static Origin-Destination (OD)-estimation from time series turnstile measurements of nodal in- and outflows in large-scale networked transit systems, where OD-information cannot be obtained from smart-card data, for instance. The approach finds posterior distribution estimates of the OD-coefficients, which describe the relative proportions of passengers travelling between origin and destination locations, via a Hamiltonian Monte Carlo sampling procedure. We suggest two different inference model formulations: the instantaneous-balance and average-delay model. We discuss both models' sensitivity to various count observation properties, and establish that the average-delay model is generally more robust in determining the coefficient posteriors. The instantaneous-balance model, however, requires lower resolution count observations and produces comparably accurate estimates as the average-delay model, pending that count observations are only moderately interfered by trend fluctuations or the truncation of the observation window, and sufficient number of dispersed data records are available. We demonstrate that the Bayesian posterior distribution estimates provide quantifiable measures of the estimation uncertainty and prediction quality of the model, whereas the point estimates obtained from an alternative constrained quadratic programming optimisation approach forego uncertainty quantification since they only provide the residual errors between the predictions and observations. Moreover, the Bayesian approach proves more robust in scaling to high-dimensional underdetermined problems, as in our specific example of the New York City (NYC) subway system. In contrast, the point estimate optimisation approach returns OD-estimates that assign large proportions of trips to few or single dedicated destinations which is incoherent with the intuitive problem assumptions. The Bayesian instantaneous-balance OD-coefficient posteriors are determined for three different times of day, based on several years of entry and exit count observations recorded at station turnstiles across the NYC subway network. The average-delay model, however, proves intractable on the real-world test scenario, given its computational time complexity and the incompleteness as well as coarseness of the turnstile records.

## 1. Introduction

The assessment, planning, and operation of transport systems relies heavily on measurements and estimates of travel demand rates between different locations of an investigated area or network. Research on Origin-Destination (OD)-estimation aims to reliably predict these travel demand rates. Due to its complexity regarding the scarcity or resolution of observable data, the number of unknown parameters, and continual technology changes, it has remained a several decade-long active field of research.

### 1.1 Brief overview of OD-estimation approaches

Considerable efforts have been devoted to OD-estimation, with notable works starting with the early beginnings of gravity models (Casey Jr, 1955; Reilly, 1931) and eventually followed by travel surveys (Ben-Akiva et al., 1985; Kuwahara and Sullivan, 1987), or inference and system identification approaches based on flow observations combined with assignment models (Bell, 1991; Cascetta, 1984; Hazelton, 2000; Maher, 1983; Nguyen et al., 1988; Robillard, 1975; Tesselkin and Khabarov, 2017; Toledo and Kolechkina, 2013; Van Zuylen and Willumsen, 1980; Yang, 1995). Recent approaches include the use of mobile phone data (Bonnel et al., 2018; Tolouei et al., 2017) or automated fare collection data (i.e., smart-cards) in transit networks (Alsger, 2017; Ji et al., 2015; Zhao et al., 2007). Generally, OD-estimation approaches are distinguished into road traffic and transit system OD-estimation, and further distilled into static and dynamic estimation methods. We refer to Antoniou et al. (2016), Cascetta et al. (2013), and Nuzzolo and Crisalli (2001) for thorough reviews of OD-estimation approaches. This paper focuses on static, inference-based OD-estimation in transit systems.

Many modern transit systems' smart-card readers have set a new paradigm for OD-estimation, because every trip record generated from a smart-card "tap-in and tap-out" sequence includes OD-information and timestamps.

Consequently, a number of recent works on transit network OD-estimation, transit assignment, and travel behaviour rely on smart-card data (Espinoza et al., 2018; Sun et al., 2015; Zhong et al., 2016). However, smart-card systems are not yet fully widespread and privacy and confidentially policies have inhibited holders of smart-card information from publicly disclosing these data. OD-estimation in transit networks thus remains a challenging research objective.

## 1.2 OD-estimation approaches relevant to this work

Most OD-matrix inference and closely related network tomography rely on measurements of the network flows between nodes, such as traffic volumes on links or the number of travellers on-board transit vehicles (Hazelton, 2010; Lawrence et al., 2006; Li and Cassidy, 2007; Nuzzolo and Crisalli, 2001; Tanaka et al., 2016; Vardi, 1996). These approaches can be distinguished into the reconstruction of actual (i.e., realized) OD trips and the estimation of the mean OD-trip rates, and typically rely on an assignment model taking into account either single or multiple route alternatives between each OD-pair (Hazelton, 2001a; Lo et al., 1996). By contrast, in many high-capacity metro networks, link flows between nodes are unknown and only the flow entering and leaving the network are measured at the stations in terms of the number of travellers who access the network at origin stations (i.e., inbound flow) and the number travellers who egress from the network at destination stations (i.e., outbound flow). Consequently, both the network link flows and the latent OD-matrix that couples the origin and destination counts are unknown. Estimates of the expected travel times, however, are typically available, with timetable information as the most approximate estimate of actual travel times.

Crucial to our problem is that we aim to estimate the OD-matrix from time series of the in- and outflow counts at every station of a transit network (e.g., daily measurements of in- and outflow within a defined observation window over a period of multiple years), without prior knowledge of an initial OD-matrix. This problem is different to a (Bayesian) update of a prior OD-matrix given a single observation of link volumes (Hazelton, 2001b; Li, 2005; Maher, 1983; Tebaldi and West, 1998). Estimating the OD-matrix using time series of passenger counts at origin and destination stations is related to the estimation of highway OD-matrices by Nihan and Davis (1987), Gajewski et al. (2002), or Park et al. (2008), and to the estimation of turning movements at road intersections by Cremer and Keller (1987). Nihan and Davis (1987) rely on a recursive method to solve the estimation problem's system of equations. However, their method requires domain-specific and experienced fine-tuning of algorithm parameters to efficiently incorporate critical sum-to-one and non-negativity constraints. Cremer and Keller (1987) propose a cross-correlation-based estimator, a least-squares estimator, or a Kalman filtering approach, that each partly relax these constraints and require additional normalization steps or approximations to enforce the constraints after a preliminary solution is found. However, these approximations and normalizations interfere with the reliability of the estimates (next to convergence issues of the Kalman filter), especially for networks with more than several tens of nodes. Cremer and Keller (1987) also propose a constrained optimization formulation that explicitly includes the sum-to-one and non-negativity constraints. Even so, optimization methods typically only perform well in low-dimensional, convex, well-defined problems, whereas realistic transit networks consist of hundreds of nodes and thus hundreds of thousands of unknown parameters. Moreover, Gajewski et al. (2002) propose an $L_2$ estimation ($L_2E$) formulation that minimizes the integrated squared error between the target and estimated densities of traffic counts as the goodness-of-fit criterion. Their formulation accounts for the non-negative sum-to-one constraints and is designed to be robust to outliers in the measured traffic volumes. However, it requires estimates of error variances that may not be trivially obtained (Park et al., 2008). In addition, the three so far discussed works, as well as most other OD-estimation methods, produce point estimates, in spite of inherently uncertain processes regarding modelling error and the real-world variability of the modelled systems. Hence, related parameters should not be considered deterministic, especially in high-dimensional problems. Park et al. (2008) thus propose a Bayesian inference formulation solved via Markov Chain Monte Carlo (MCMC) to account for estimation uncertainty, while also imputing missing values and being insensitive to outliers in measured traffic volumes. However, Park et al. (2008) do not account for prior sum-to-one constraints (only non-negativity), which is found to be critical in the applications considered here, with high-dimensional parameter spaces and strongly underdetermined problems. In fact, in all four of the discussed works, the models and methods are only demonstrated on a limited number of up to 10 count locations and exclusively deal with overdetermined problems (i.e., the number of observations exceeds the number of unknown model parameters). Instead, this work addresses OD-estimation problems that are underdetermined, with for instance in the order of $10^5$ parameters and $10^3$ observations. It is therefore critical to explicitly include non-negative sum-to-one parameter constraints that drastically and favourably reduce the solution domain. Moreover, works so far have not studied the sensitivity of estimation accuracy and precision to the properties of the count observations and measurement parameters. At last, the four aforementioned works all consider a single simplified model that neglects travel time between locations of the network. Benchmarking against a more detailed model in order to assess how strongly the neglected travel time assumption affects the estimation results has not been addressed.

## 1.3 This work's contributions

In light of the above overview, most approaches to OD-estimation require network-internal link flow measurements that are not readily available for the cases studied here. Moreover, they suffer from the curse of dimensionality and/or do not explicitly treat estimation uncertainty related to model assumptions and observation variability. To comprehensively address these challenges, we propose two Bayesian inference models with different levels of complexity regarding the underlying network flow assumptions. Both models identify OD-estimates in large-scale urban transit networks based on time series measurements of the in- and outflow counts at stations. We term these two models the instantaneous-balance model and the average-delay model, where the instantaneous-balance model is subject to more approximate assumptions and constraints. They both formulate joint posterior distributions of the OD-coefficients that constitute the OD-matrix. These OD-coefficients determine the relative flow proportions bound for every destination, given an origin. Since the model solutions are analytically intractable, posterior distribution estimates are determined from a Hamiltonian Monte Carlo sampling approach. We demonstrate and compare the performance of the two models on test networks, and highlight the required preconditions for these models to generate reliable (i.e., accurate and precise) OD-coefficient estimates. Moreover, this comparison serves to evaluate the degree to which the simplified instantaneous-balance model is able to match the more detailed average-delay model results. At last, we showcase the methods on the New York City subway system, with several years of publicly available real-world turnstile count data. The proposed approach offers a quantification of the parameter uncertainty and delivers estimates that are at least as accurate as existing optimization-based results for small-scale test networks, while not suffering from the curse of dimensionality as the size of the network increases.

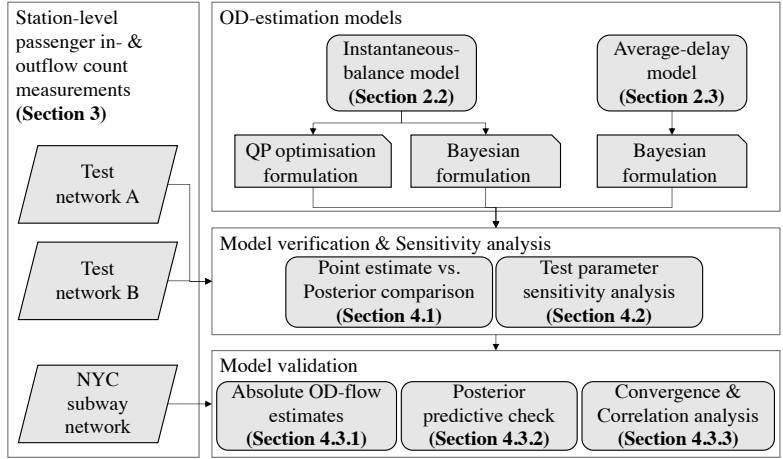

**Figure 1.** Workflow of the Bayesian OD-estimation model development, verification, and validation steps.

## 2. Approach

Fig. 1 shows the workflow of this analysis. The starting point are time series of the in- and outflow counts of travellers at each station in a transit network. Since these data do not explicitly express the number of passengers travelling between the origin and destination locations, we develop two models to infer the latent OD-matrix. These models exploit that multiple measurements (i.e., time series) of nodal in- and outflow counts are recorded and, if possible, incorporate additional network information, namely, expected travel times that can for instance be obtained from service timetables. The following sections delineate these two model formulations that are developed to identify the elements of the OD-matrix, i.e., the "instantaneous-balance model" and the "average-delay model".

For both model formulations, the aim is to identify the parameter values of the OD-matrix from entry and exit count observations at $S$ number of stations (also called nodes) in a fully-connected network. The OD-matrix $A_{S \times S}$ contains the OD-coefficients $\alpha_{ij}$ for each origin node $i \in \{1, 2, \ldots, S\}$ and destination node $j \in \{1, 2, \ldots, S\}$,

$$A = \begin{bmatrix} 0 & \alpha_{12} & \cdots & \alpha_{1S} \\ \alpha_{21} & \ddots & \ddots & \vdots \\ \vdots & \ddots & \ddots & \alpha_{(S-1)S} \\ \alpha_{S1} & \cdots & \alpha_{S(S-2)} & 0 \end{bmatrix}_{S \times S}, \qquad [1]$$

with

$$\sum_{j=1}^{S} \alpha_{ij} = 1; \qquad \alpha_{ij} \geq 0; \qquad \alpha_{ij} = 0 \; \forall \; i = j.$$

The diagonal in Eq. (1) is zero, as no self-loops are permitted, i.e., none of the inflow leaves the network at its origin node. The OD-coefficients may be interpreted from two viewpoints; we can view OD-coefficients as the relative trip proportions of travellers entering at specific origins and bound for specific destinations; as such, they are also referred to as split coefficients (Cremer and Keller, 1987; Nihan and Davis, 1987) or split proportions (Gajewski et al., 2002; Park et al., 2008). Alternatively, OD-coefficients can be viewed as the conditional choice probabilities of selecting destination $j$, given that a traveller accesses origin $i$. We will use both viewpoints interchangeably, minding that they will converge in the limit of a large number of passengers.

## 2.1 Network assumptions

We rest our analysis on several key assumptions regarding the structure of the network and the observation sequence, that are in line with the assumptions by Gajewski et al. (2002) and Park et al. (2008):

1) *Observability:* The network consists of links and nodes (i.e., edges and vertices). All points of the network where access and egress is possible are observable and passenger counts can be measured over steady state and periodic observation windows.
2) *Steady state:* The OD-matrix and system operations stay constant during a single observation window (e.g., the OD-matrix remains constant during the 8:00 to 9:00 AM observation window).
3) *Periodicity:* The OD-matrix repeats periodically along with recurrent system operations such as schedules and rolling stock movement during every observation (e.g., week-day train schedules remain unchanged over the entire recording span of passenger in- and outflows).
4) *Mass conservation:* The total inflow has to equal the total outflow across the network, i.e., every passenger who accesses the network has to eventually egress from the network – both the access event and the egress event need to be measurable.

Both the instantaneous-balance and the average delay model formulation adhere to the above assumptions. However, the instantaneous-balance model relaxes that passengers need time to travel between stations in the network, that is, mass conservation is achieved instantaneously such that over any chosen observation time window the total number of entering passengers balances with the total number of exiting passengers. The average-delay model therefore extends the list of assumptions, by including a fifth item,

5) *Expected travel time:* Between every OD-pair exists an expected (measurable) travel time. When multiple route options exist between an OD-pair, the expected travel time is the mean travel time over all route options falling within a defined time margin from the earliest arrival (shortest) path.

As a result, the average-delay model requires fine grained, time-resolved observational data to construct a mapping between the entry and exit count observations. We present a detailed comparison of the instantaneous-balance and average-delay model.

## 2.2 Instantaneous-balance model

A simple model linking entry and exit count observations is the linear relation

$$\underset{N\times S}{Y} = \underset{N\times S}{X} \cdot \underset{S\times S}{A} + \underset{N\times S}{\epsilon}, \qquad [2]$$

where $Y$ is a matrix of size $N \times S$ and contains the number of exiting passengers at all $S$ stations during $N$ observations. The matrix $X$ contains the corresponding entry count observations for all stations during the same observations. The second right-hand side term in Eq. (2) captures the errors $\epsilon$ due to model deficiencies and measurement noise. The inverse problem associated with Eq. (2) is solved by Cremer and Keller (1987) for the turning movements at intersections through a point-estimate-based constrained optimisation technique. By way of comparing the proposed Bayesian formulations and solution approaches against the point estimate formulation, Appendix A similarly expresses this problem as a constrained quadratic optimisation programme. The optimisation objective function includes an intercept variable and additional regularisation terms that improve the OD-coefficient estimates, and are not included into the original constrained optimization by Cremer and Keller (1987). These regularisation terms are detailed in the context of the proposed model in Section 2.2.2.

### 2.2.1 The instantaneous-balance Bayesian model

When solving the quadratic optimisation programme in Eq. (A.1) of Appendix A, the returned solution gives point estimates of the OD-coefficients. These point estimates do not represent the inherent estimation uncertainty and are not robust for high-dimensional instances. In order to address these limitations, we express the model in Eq. (2) as a Bayesian inference model. The OD-split coefficient parameters $\alpha_{ij}$ thus are treated as random variables. Moreover, the model includes auxiliary parameters, namely, the scale parameter $\sigma_{y,j}$ and intercept $r_j$ to account for measurement and model error. Following Bayes' rule, the posterior density for the unknown OD-split coefficients $\alpha_{ij}$ (and parameters $\sigma_{y,j}$ and $r_j$), denoted by $\pi(\alpha, \sigma, r|\mathcal{D})$, is expressed in terms of the likelihood $p(\mathcal{D}|\alpha, \sigma, r)$ and prior $\pi(\alpha, \sigma, r)$ according to

$$\pi(\alpha,\sigma,r|\mathcal{D}) = \frac{p(\mathcal{D}|\alpha,\sigma,r)\,\pi(\alpha,\sigma,r)}{\iiint p(\mathcal{D}|\alpha,\sigma,r)\,\pi(\alpha,\sigma,r)d\alpha d\sigma dr} \propto p(\mathcal{D}|\alpha,\sigma,r)\,\pi(\alpha,\sigma,r)\,. \qquad [3]$$

The evidence $\mathcal{D}$ are the exit count observations $y_j^{(n)}$ at station $j \in \{1, 2, \dots, S\}$ and during observation instance $n \in \{1, 2, \dots, N\}$ – In this model, the exit counts are considered observational data, whereas entry counts $x_i^{(n)}$ are model constants with $i \in \{1, 2, \dots, S\}$. The likelihood is specified according to

$$p(\mathcal{D}|\alpha,\sigma) = \prod_{j=1}^{S} \prod_{n=1}^{N} p\!\left(y_j^{(n)} \big| \mu_{y,j}^{(n)}, \sigma_{y,j}\right), \qquad [4]$$

where every observation of the number of exiting passengers $y_j^{(n)}$ is a realisation of a truncated normal distribution (lower truncated at zero), $N^+$, with location parameter $\mu_{y,j}^{(n)}$ and observation-invariant scale parameter $\sigma_{y,j}$, such that

$$y_j^{(n)} \big| \mu_{y,j}^{(n)}, \sigma_{y,j} \sim N^+\!\left(\mu_{y,j}^{(n)}, \sigma_{y,j}\right), \qquad [5]$$

with

$$\mu_{y,j}^{(n)}(\alpha, x) = \sum_{\substack{i=1 \\ i \neq j}}^{S} \alpha_{ij} x_i^{(n)} + r_j\,. \qquad [6]$$

The truncated normal distribution reflects that the exit counts $y_j^{(n)}$ are constrained to be non-negative. The expressions in Eq. (5) and Eq. (6) comply with the model in Eq. (2), where the error term is absorbed into the scale parameter $\sigma_{y,j}$ and intercept $r_j$. While the scale parameter accounts for uncertainties that are related to random errors, the intercept accounts for any biasing errors imposed by the model.

Whereas the likelihood captures the assumed model and weighs it against the evidence given by the observed data, the prior introduces any domain knowledge or information about the model parameters. The prior is specified according to

$$\pi(\alpha,\sigma,r) = \prod_{i=1}^{S} p(\alpha_{i\bullet}) \prod_{j=1}^{S} p(\sigma_{y,j}) \prod_{j=1}^{S} p(r_j)\,. \qquad [7]$$

The $i\bullet$ notation indicates the row-wise simplex parameterization for $\alpha_{ij}$, i.e., the non-negative, sum-to-one OD-coefficient constraint according to Eq. (1). As comparison, Park et al. (2008) define a truncated normal distribution for the OD-coefficients. In addition, we place a symmetric Dirichlet hyperprior on each row of the OD-matrix, according to

$$\underset{1 \times (S-1)}{\boldsymbol{\alpha}_{i\bullet}} \sim Dir(c\mathbf{1})\,, \qquad [8]$$

where $\mathbf{1}$ is a vector of ones with size $1 \times (S-1)$, and $c$ is a concentration parameter. The symmetric Dirichlet hyperprior in Eq. (8) maintains the simplex constraint and includes additional knowledge about the clustering of OD-coefficient values. By adjusting the concentration parameter in Eq. (8) away from 1, the prior biases towards either sparse ($c < 1$) or dense ($c > 1$) OD-coefficient matrix. Appendix B shows an example of a two-dimensional simplex and the influence of the concentration parameter. Unless domain expertise strongly suggests either a more sparse or dense OD-matrix, $c$ should be set to 1. Alternatively, the Dirichlet hyperprior in Eq. (8) can be parameterized as $\boldsymbol{\alpha}_{i\bullet} \sim Dir(\boldsymbol{c}_i)$, where $\boldsymbol{c}_i$ is a vector of concentration parameters $c_{ij}$ for each destination $j$. Enforcing zero constraints similar to related works by Gajewski et al. (2002) and Park et al. (2008) for infeasible OD-coefficients ($\alpha_{ij} = 0$) is then achieved by setting the corresponding concentration parameters to zero ($c_{ij} = 0$).

The scale parameters $\sigma_{y,j}$ as well as the intercepts $r_j$ are given improper flat (uniform) priors according to

$$\sigma_{y,j} \sim U(0, \infty)\,, \qquad [9]$$

$$r_j \sim U(-\infty, \infty).$$

We assume that the likelihood model is informative enough to prevent $\sigma_{y,j}$ and $r_j$ from diverging towards infinity.

### 2.2.2 Model regularisation

The intercept $r_j$ in Eq. (6) corrects for the average biasing error between the linear model and the observable range of in- and outflow counts. However, an inadvertent consequence may be that the intercept model overfits the observed data, capturing the observable range overly close and performing poorly when predicting on unseen data. Consequently, we extend the instantaneous-balance model by two regularisation terms. Firstly, we include a minimal-bias component that aims to minimise the squared sum of the intercepts $r_j$, by defining an additional probability density according to

$$p\left(\sum (r_j)^2\right) = \frac{1}{C}\exp\left(-\sum (r_j)^2\right) \propto \exp\left(-\sum (r_j)^2\right). \qquad [10]$$

We can omit the normalizing factor $C^{-1}$, by assuming that it is constant and does not depend on $r_j$.

Secondly, we include an expected value component, that tests the instantaneous-balance assumption against the mean entry and exit count observations. This component enters the likelihood model, and is defined according to

$$\bar{y}_j | \mu_{\bar{y},j}, \sigma_{y,j} \sim N^+(\mu_{\bar{y},j}, \sigma_{y,j}), \qquad [11]$$

with

$$\mu_{\bar{y},j}(\alpha) = \sum_{\substack{i=1 \\ i \neq j}}^{S} \alpha_{ij}\, \bar{x}_i, \qquad [12]$$

where $\mathbb{E}[X_i] \approx \bar{x}_i = \frac{1}{N}\sum_{n=1}^{N} x_i^{(n)}$ and $\mathbb{E}[Y_j] \approx \bar{y}_j = \frac{1}{N}\sum_{n=1}^{N} y_j^{(n)}$ are the sample means of the entry and exit count observations. Eq. (12) does not include an intercept.

By including the regularisation terms, we aim to control the generalizability (i.e., extrapolation outside of the observable range) of model predictions and avoid overfitting of the observed data. The minimal-bias and expected value expressions in Eq. (10) and (11) propose that the missing-intercept model (i.e., the uncorrected instantaneous-balance assumption) is a generalisable approximation of the true underlying processes and thus valid both within and beyond the observable range, whereas the likelihood model in Eq. (5) and (6) weighs in the proposal that the intercept model (i.e., the corrected instantaneous-balance assumption) captures more closely the observable range. The result is a trade-off between these two proposals that reflects positively in finding generalisable parameter estimates.

## 2.3 The average-delay model

The instantaneous-balance model neglects the time lags involved in travelling between origins and destinations. This assumption can give rise to significant estimation errors of the OD-matrix. In order to overcome this limitation, we propose the average-delay model, where delay refers to the travel time or lag between an origin and destination.

### 2.3.1 Formal model description

Fig. 2 shows diagrams of the passenger flow process in the average-delay model. Observations are recorded periodically, e.g. daily, (Fig. 2, top panel). Every observation $n \in \{1, 2, \ldots, N\}$ considers a departure window and an arrival window (Fig. 2, bottom panel). The departure window starts at $t = t_0$ and comprises earlier times than the arrival window, which starts at $t = t_1$. Both end at the same time $t = T$. The time gap $t_1 - t_0$ between the start of the departure window and the start of the arrival window is chosen as the longest travel time between any two locations in the network. This ensures that all exit counts observed within the arrival window have a matching entry count within the departure window. Alternatively, the time gap can be selected based on domain knowledge. For instance, the maximum planned travel time of any traveller may be assumed shorter than 1 hour and thus the time gap would be set to 1 hour. The departure and arrival window are discretised into time bins.

We determine the average travel time between every OD-pair (e.g., from timetable information) and construct an OD-assignment matrix $H_{K \times S}$, whose elements are the number of entering passengers at origin $i \in \{1, 2, \ldots, S\}$ during observation $n$ and time bin $t \in \{1, 2, \ldots, T\}$. A share of these entering passengers exits at destination $j$ with a delay of $t + \Delta \bar{t}_{ij}$, and same observation $n$. This share is given by the OD-coefficient $\alpha_{ij}$, that is, the proportion of entering passengers at $i$ that are bound for $j$. The average travel time $\Delta \bar{t}_{ij}$ is given in terms of the number of lagged time bins required to travel between the two stations, and is assumed as given either because it is measurable or can be estimated external to the OD-estimation problem. The OD-assignment matrix has $K = S \times N \times T_a$ rows, where $T_a$ is the number of arrival time bins.

Formally, the average-delay model describes the exit count observations at destination $j$, according to

$$\underset{(N \times T_a) \times 1}{\boldsymbol{y}_j} = \underset{(N \times T_a) \times S}{\boldsymbol{H}_k} \cdot \underset{S \times 1}{\boldsymbol{\alpha}_j} + \underset{(N \times T_a) \times 1}{\boldsymbol{\epsilon}} \qquad [13]$$

where $\boldsymbol{y}_j$ is a column vector of length $N \times T_a$, and contains the number of passengers exiting at station $j$ during time bin $t_a \in \{1, 2, \ldots, T_a\}$ and observation $n \in \{1, 2, \ldots, N\}$. The sub-matrix $\boldsymbol{H}_k$ contains the entry count observations at origin $i \in \{1, 2, \ldots, S\}$ during observation $n \in \{1, 2, \ldots, N\}$ at lagged time bin $(t_a - \Delta \bar{t}_{ij})$ with respect to destination $j$. The column vector $\boldsymbol{\alpha}_j$ corresponds to the $j$-th column of the OD-matrix $\boldsymbol{A}$ in Eq. (1). The second right-hand side term in Eq. (13), $\boldsymbol{\epsilon}$, accounts for the modelling and measurement errors. The element $h_{kj}$ at index $(k, j)$ of the OD-assignment matrix is given by

$$h_{kj} = \begin{cases} x_i^{(n, t_d)}, & \text{if } t_a \in \{1, 2, \ldots, T_a\}, \\ 0, & \text{otherwise,} \end{cases} \qquad [14]$$

where $x_i^{(n, t_d)}$ corresponds to the entry counts at station $i$ during observation $n$ and departure time bin $t_d \in \{1, 2, \ldots, T\}$, with $T$ being the number of departure time windows. The row index $k$ is given in terms of the origin station $i$, destination station $j$, the observation $n$, and the arrival time bin $t_a$, according to

$$k = (j - 1) \times N \times T_a + (n - 1) \times T_a + t_a . \qquad [15]$$

The arrival time bin depends on the specific OD-pair and is given by

$$t_a = t_d + \Delta \bar{t}_{ij} - t_1 + 1 . \qquad [16]$$

### 2.3.2 The average-delay Bayesian model

We express the average-delay model into a Bayesian formulation. The posterior density for the unknown parameters $\alpha_{ij}$ is given by

$$\pi(\alpha, \sigma | \mathcal{D}) \propto p(\mathcal{D} | \alpha, \sigma) \, \pi(\alpha, \sigma) . \qquad [17]$$

where the evidence $\mathcal{D}$ are the exit count observations $y_j^{(n, t_a)}$ at station $j$, during observation $n$ and time bin $t_a$. Moreover, the model includes only the auxiliary scale parameters $\sigma_{y,j}$ to account for measurement and modelling errors. Given that the average-delay model accounts more closely for travel time and truncates any passenger arrivals outside the arrival window, we omit the intercepts $r_j$, that capture the average bias errors in the instantaneous-balance model. We define the likelihood as

$$y_j^{(n, t_a)} | \mu_{y,j}^{(n, t_a)}, \sigma_{y,j} \sim N^+ \left( \mu_{y,j}^{(n, t_a)}, \sigma_{y,j} \right) , \qquad [18]$$

with

$$\mu_{y,j}^{(n, t_a)}(\boldsymbol{\alpha}, \boldsymbol{H}) = \sum_{\substack{i=1 \\ i \neq j}}^{S} \alpha_{ij} h_{ki} . \qquad [19]$$

$N^+$ in Eq. (18) denotes the truncated normal distribution. The prior is specified according to

$$\pi(\alpha, \sigma) = \prod_i^S p(\alpha_{i \bullet}) \prod_j^S p(\sigma_{y,j}) . \qquad [20]$$

The OD-split coefficients $\alpha_{ij}$ are constrained as simplexes for each origin station $i$ and we place the symmetric Dirichlet hyperprior in Eq. (8). The observation- and time bin-invariant scale parameters $\sigma_{y,j}$ are constrained by improper uniform priors,

$$\sigma_{y,j} \sim U(0, \infty) \,. \qquad [21]$$

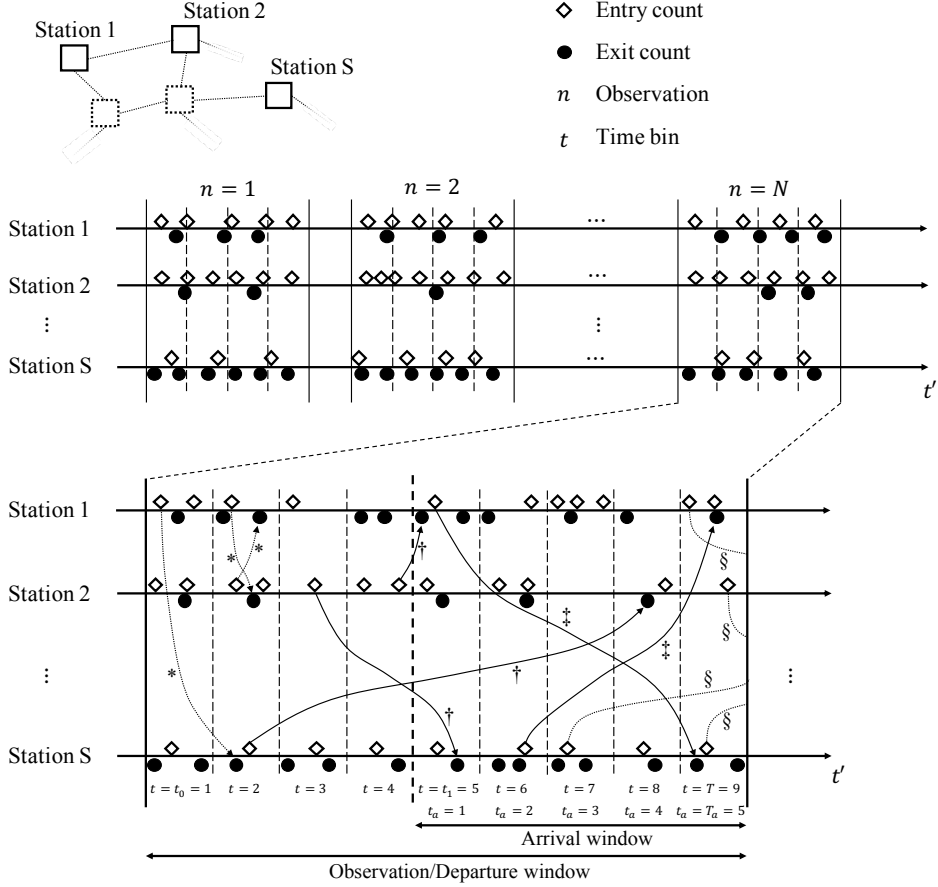

**Figure 2.** *Top panel:* Schematic of the assumed average-delay observation sequence for a network consisting of $S$ stations. The stations are connected through a fully-connected network with known timetable information. Entry and exit counts are collected at every station within $N$ periodical observation windows. Every observation window is subdivided into time bins, starting with time bin $t = t_0$ and ending at $t = T$. System conditions and the underlying latent origin-destination matrix are assumed constant across all observations. One entry count or exit count corresponds to one passenger entering or exiting at a station, respectively. *Bottom panel:* The assumed access, travel time, and egress process for one observation window of the average-delay model with $T = 9$ time bins. The departure window starts at $t = t_0 = 1$. The arrival window starts at $t = t_1$, $(t_1 - t_0)$ time bins after the departure window to account for passenger trips departing prior to and arriving during the arrival window. *Legend:* The arrows indicate example trips of passengers; *The passenger departed and arrives pior to the start of the arrival window; †The passenger departed prior to the arrival window and arrives during the arrival window; ‡The passenger departed and arrives during the arrival window; §The passenger arrival time at the destination station occurs after the observation window.

## 2.4 OD-Matrix Estimate via Markov Chain Monte Carlo

The point estimates, that are for instance obtained from the optimization model in Appendix A, do not express the prevalent parameter uncertainty due to modelling and measurement errors. In contrast, the Bayesian approach propagates the uncertainty regarding model assumptions and measurements into the posterior distribution estimates of the OD-coefficients. These posterior distribution estimates can hence be used to quantify each parameter's expectation, that is, "the weighted average" over its (marginal) posterior distribution, as detailed in Appendix C.

The OD-estimation problem proposed in this paper aims to find estimates for $S \times (S - 1)$ OD-coefficients and $S$ scale parameters, with additional $S$ intercepts for the instantaneous balance model. Given the multidimensional parameter space and constraints, the models in Eq. (3) and (17) are analytically intractable, and prevent from obtaining a closed form expression of the OD-coefficients' posterior distribution. Moreover, the high-dimensional parameter space makes numerical approximation of the normalization integral associated with Bayes' rule practically infeasible. Markov Chain Monte Carlo (MCMC) approaches, however, are able to find sampling estimates of the posterior distribution (Geyer, 2011). In the context of MCMC, the targeted posterior distribution is also known as the equilibrium distribution.

We thus resort to MCMC sampling to estimate the multidimensional posterior distribution $\pi(\alpha, \sigma, r | \mathcal{D})$ of the OD-matrix coefficients and auxiliary parameters. Traditional MCMC samplers like the Metropolis-Hastings algorithm struggle to converge and sample from the equilibrium distribution of high-dimensional parameter spaces, since they are based on random walk processes. The particularly large number of OD-coefficients thus poses a problem. Hamiltonian Monte Carlo (HMC), however, has proven to overcome this limitation by taking targeted and tuned steps through the posterior distribution based on derivative information of the target distribution and numerical integration in a transformed parameter space (Duane et al., 1987; Neal, 2011, 1996). The result is a significant improvement in sampling efficiency. We apply the HMC variant known as NUTS (No-U-Turn Sampler) (Hoffman and Gelman, 2014), which includes tuning of the algorithm parameters prior to sampling from the target distribution. At last, the MCMC sampling approach is also the reason why the normalizing constant in Eq. (10) can be omitted, as it cancels out during the derivative and Metropolis acceptance step.

## 3. Generative processes and test networks

The capability of the instantaneous-balance and average-delay model in estimating the latent OD-matrix coefficients of transit systems is demonstrated on two synthetic datasets and on one real-world dataset. Test network A encompasses the synthetic dataset to test whether the inverse problem tackled by the instantaneous-balance inference model is solvable. It entails a minimum level of network information, i.e., it only provides daily entry and exit counts, and consists of 15 stations with its link structure unknown. The generative process for test network A is detailed in Appendix E. Real-world data do not entirely fulfil the assumptions of the instantaneous-balance model, which is, therefore, subject to modelling error. For instance, we expect discrepancies between the observed and predicted exit counts, since the instantaneous-balance model omits travel time information between stations. Therefore, test network B enforces realistic assumptions regarding the generative processes associated with the observational data. Foremost, several path alternatives with varying travel times exist between each OD-pair. In addition, the number of entering and exiting passengers within the fixed and truncated observation windows are unbalanced. Test network B starts from simulated daily entry counts and generates exit counts via the path choice and associated travel delay models. The generative process for test network B is detailed in Appendix F. It is based on a subcomponent and timetable data of the New York City (NYC) subway (Metropolitan Transport Authority, 2019a), consisting of 35 stations, 8 lines, and 39 transfer linkways (i.e., footpaths) shown in Fig. F.1. Test network B encompasses the synthetic dataset to compare the instantaneous-balance model to the higher resolution average-delay inference model. It is therefore used to assess the ramifications of the instantaneous-balance model assumptions, when tested on realistic data and to benchmark the gap with respect to the higher resolution average-delay model.

Next to generic model testing, the synthetic datasets are used to demonstrate model robustness against the properties of the underlying generative processes and chosen observation horizons. For test network A, the variance of the entry and exit count observations is modulated through an adjustable dispersion parameter $\phi$. Test network B, additionally includes a trend scale $\eta$ to control within-day entry count fluctuations, and an observation window width $w$ to adjust the daily observation horizon.

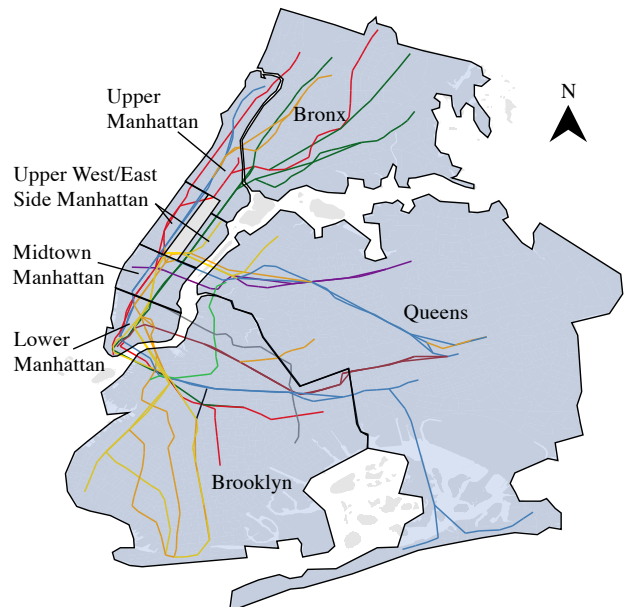

**Figure 3.** The New York City subway network and city boroughs (Metropolitan Transport Authority, 2019b). The borough of Manhattan is divided into Lower, Midtown, and Upper Manhattan. The fifth borough, Staten Island, is not shown on this map.

### 3.1 Real-world dataset

In addition to synthetic datasets, we aim to take a step towards validating the proposed models on real-world datasets, i.e. the NYC subway network shown in Fig. 3. Starting in April 2010, the MTA publishes turnstile entry and exit counts for every station (Metropolitan Transport Authority, 2019a). The count data are originally collected in 4-hour intervals and recorded together with a turnstile identifier. Every station has multiple turnstiles. We conservatively up-sampled the counts to a higher-resolution time granularity. The up-sampling method is based on a monotonic rational quadratic spline interpolation (Delbourgo and Gregory, 1983) of the cumulative, and thus monotonically increasing, 4-hour interval count values. The up-sampled cumulative count values are differenced to obtain the synthetic higher-resolution count data. The $C^2$-continuity of the interpolation ensures that the differenced interpolated values remain smooth. We chose a 5-minute upsampling interval. These upsampled counts are conservative in that they sum to the original 4-hour interval count data. In addition, we apply an auto-encoder data filtering approach to correct outliers and missing data that are interspersed throughout the real-world dataset. Details about the up-sampling approach and subsequent filtering procedure for outlier detection and missing data imputation are given in Appendix G.

Furthermore, the observed total exit counts of the NYC subway system are consistently lower than the observed total entry counts. For instance, the average total number of entering passengers is 5.9 million versus 4.3 million exiting passengers, when summed over a full daily cycle. This is presumably due to passengers using emergency exits instead of the official turnstile gates. In order to comply with the mass conservation assumption, we consequently balanced the daily in- and outflow, by distributing the excess inflow proportionally across stations as given by the procedure in Appendix D. Assuming that the entry counts follow Negative Binomial distributions for every station, we determine the equivalent dispersion parameter values, given the sample mean and variance of the entry count data. The dispersion parameter concentrates near values of order $10^2$, indicating mostly overdispersed count data.

These pre-processing steps result in time series of the number of travellers accessing and egressing at each station in the NYC subway system in 5-minute intervals over a time span of approximately 8 years (April 2010 to August 2018). In this analysis, however, we only consider weekday counts and the months from April until October. Moreover, the counts are aggregated over pre-defined observation windows, as detailed in Section 4.3. The purpose of constraining the data records in this way is to comply with the models' assumptions regarding recurrent system conditions and unchanged OD-demand. The resulting number of observations is equal to $N = 1315$ records of the in- and outflow counts of passengers at each of the $S = 471$ stations of the NYC subway system[1].

### 3.2 Accuracy and precision of OD-coefficient estimates

Since the true OD-matrices of test network A and B are known, we can quantify the accuracy of the OD-coefficient estimates. We define the Bayesian posterior accuracy in terms of the mean-squared-error (MSE) between the posterior means $\hat{\mu}_{\alpha_{ij}}$ and the OD-coefficients' true values $\tilde{\alpha}_{ij}$. In addition, we define the accuracy of the QP optimisation solution in terms of the MSE between the point estimates $\hat{\alpha}^*_{ij}$ and the true values $\tilde{\alpha}_{ij}$, such that

$$\text{MSE-MCMC} = \frac{1}{S^2} \sum_{i=1}^{S} \sum_{j=1}^{S} \left( \hat{\mu}_{\alpha_{ij}} - \tilde{\alpha}_{ij} \right)^2,$$
$$\text{MSE-QP} = \frac{1}{S^2} \sum_{i=1}^{S} \sum_{j=1}^{S} \left( \hat{\alpha}^*_{ij} - \tilde{\alpha}_{ij} \right)^2. \qquad [22]$$

The highest-posterior-density (HPD) interval of the Bayesian OD-coefficient posterior is a measure of the estimates' uncertainty. We define the aggregate uncertainty statistic, which we will refer to as the estimates' precision, in terms of the mean 95% HPD interval across all coefficient posterior estimates,

$$\hat{\mu}_{\text{HPD}} = \frac{1}{S^2} \sum_{i=1}^{S} \sum_{j=1}^{S} \text{HPD}(\alpha_{ij})_{95\%} . \qquad [23]$$

---

[1] In fact, there are currently 472 stations in the NYC subway system. This study is based on data prior to the re-opening of the station Cortlandt Street in 2018, after being rebuilt in the aftermath of the 9/11 attacks.

We do not define a precision statistic for the optimisation results because they are point estimates. The accuracy statistic measures the estimation error with respect to the true values, whereas the precision statistic measures the estimates' posterior resolution irrespective of how far the estimates are from their true values.

## 4. Results and discussion

We encode and run the proposed models in *Stan* (Carpenter et al., 2017), which implements the NUTS algorithm. Data processing is implemented in Python 3.6. The model code and compilation of the model are executed on the Python package *pystan*. In order to satisfy the OD-coefficients' non-negative sum-to-one constraint, the coefficients are parameterised as simplexes. The definition of a simplex poses an inherent sampling difficulty, especially in high dimensions, given that any new sample in an arbitrary direction is hardly guaranteed to satisfy the simplex constraint. *Stan* implements a stick-breaking transformation to map from the $(S-1)$-dimensional simplex to an unconstrained, more sampling efficient $(S-2)$-dimensional space.

Throughout the analysis, we assume unbiased prior knowledge about the concentration of the OD-coefficients and set $c$ in Eq. (8) to 1. This results in a flat Dirichlet distribution of the OD-coefficients, i.e. the probability mass is uniformly spread over the entire domain of the multidimensional OD-coefficient simplex support. To estimate the average travel time of path alternatives between origins and destinations for the average-delay model, the travel times of all path alternatives falling within a 10-minute arrival time margin from the earliest arrival path are averaged.

### 4.1 Instantaneous-balance model verification

We consider test network A, estimate the OD-coefficients based on the instantaneous-balance model using the proposed Bayesian inference approach, and compare these estimates against the QP estimates. The instantaneous-balance model results are based on the regularized formulations of Section 2.2.2 (Appendix H provides additional detail on the comparison between the regularized and unregularized formulation results).

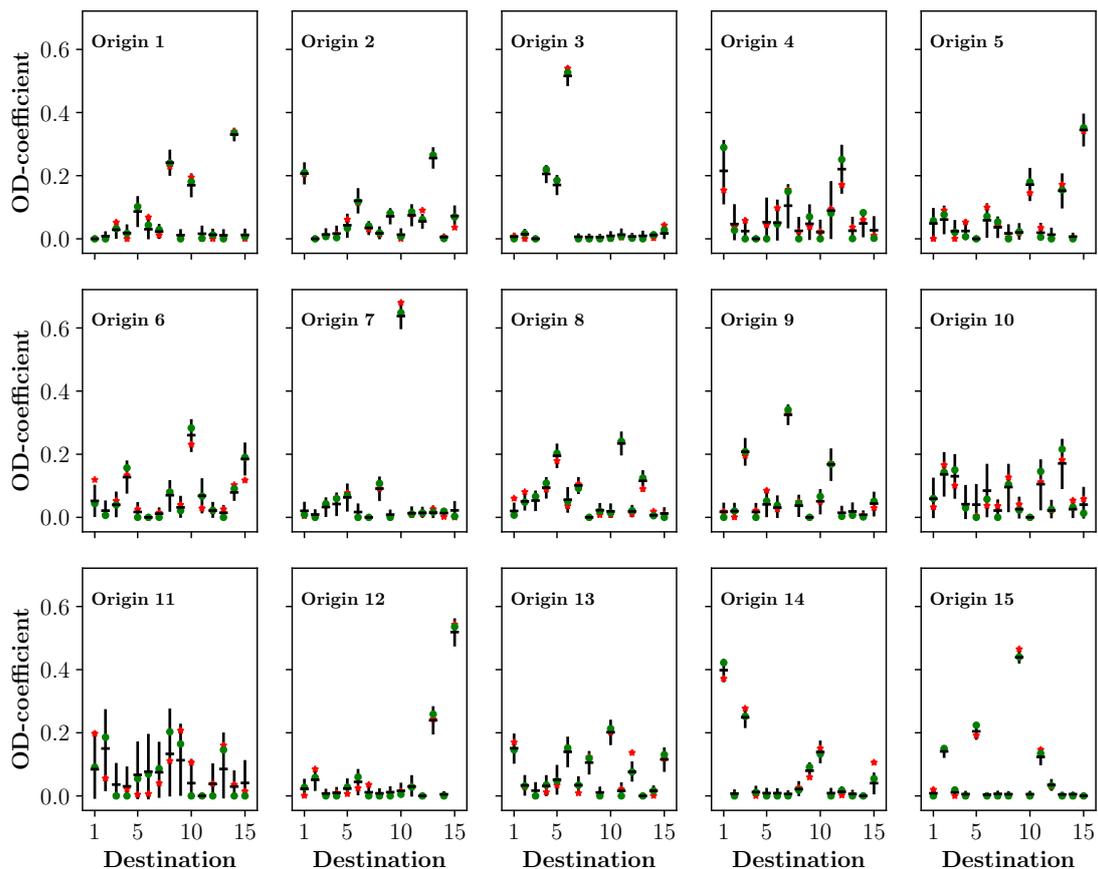

**Figure 4.** OD-coefficient estimates for a 15-node network, based on $N = 30$ entry and exit count observations at each station node. The entry count dispersion parameter of the generative process model is $\phi = 10$. The red stars mark the true OD-coefficient values, the green dots indicate the quadratic programme solution, and the horizontal dashes mark the sample means of the MCMC posterior estimate. Vertical lines indicate the 95% probability highest posterior density (HPD) intervals of the MCMC posterior estimate.

### 4.1.1 Bayesian and QP model comparison

Fig. 4 illustrates the OD-coefficient estimates according to both approaches[2], using observations from the generative process in Appendix E with $\phi = 10$ and $N = 30$. Both the QP optimisation and the Bayesian MCMC approach reliably reconstruct the latent OD-matrix from only the entry and exit count observations of test network A. However, the QP estimates do not provide a measure of the parameters' uncertainty. In contrast, the vertical bars in Fig. 4 indicate the 95% probability HPD interval, which provides uncertainty bounds of the MCMC-derived parameter estimates. Remarkably, several origin stations consistently exhibit coefficient estimates with large uncertainties (i.e., wide HPD intervals), which are correlated with small mean entry counts (as reported in Appendix E, Table E.1). This is because the exit counts at destination stations are the result of compounded OD-flows from all origin stations, and, consequently, OD-flows that originate from stations with small inflows are more difficult to discern against the more dominant contributions of high-inflow stations.

The posterior predictive distribution of the exit counts is found by marginalizing the likelihood model over the posterior distribution. We determine the posterior predictive distribution during the MCMC sampling routines, by using the sampled OD-coefficient values and substituting them into the model description in Eq. (6) and sampling from the truncated normal distribution in Eq. (5). Fig. 5 shows the posterior predictive distributions for a single observation of the exit counts at the 15 stations of test network A. The observations fall within the predictive distribution's support, indicating that the model can recover the structure of the observed data. In contrast, the QP-derived point estimates only provide a crisp value and their accuracy can only be judged by their residual error, which in this example also deviates from the observations for some stations. These deviations from the observed counts are generally due to model inadequacies and limited sample size – Here, the deviations are merely due to the limited sample size, since both the inference model and the generative process are based on the same network assumptions. Appendix H provides additional discussions on the influence of the number of observations on accuracy.

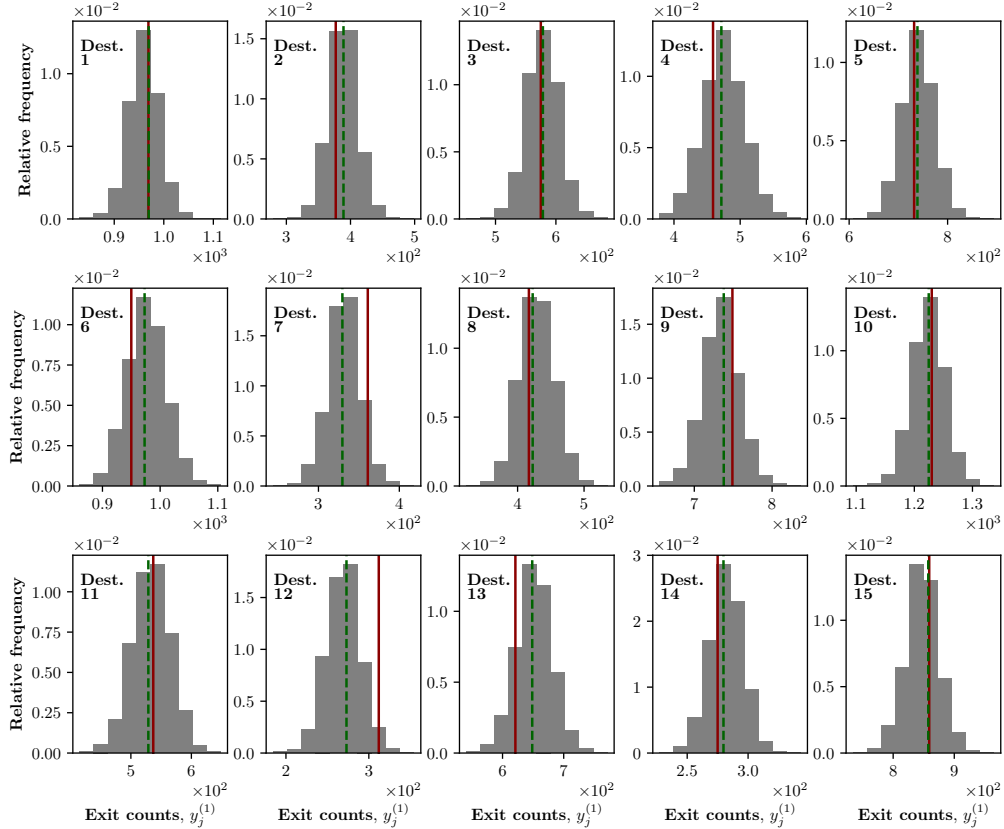

**Figure 5.** Predictive test on a single count observation under instantaneous-balance model assumptions. The red vertical line indicates the observed exit counts; The dashed green line indicates the prediction based on the QP coefficient point estimates; The greyed histogram indicates the posterior predictive distribution based on the MCMC coefficient posterior estimates.

---

[2] The MCMC sampler is configured to the default control and sampling parameters. The number of warm-up iterations is set to 1500; the number of sampling iterations is 1000. The sampler is randomly initialised on four chains, resulting in a total of 4000 posterior samples. The MCMC sampler does not exhibit any divergences and none of the tree searches saturates at the maximum tree-depth of 10. The potential scale reduction statistic (c.f., 0), $\hat{R}$, is close to 1.0 for all parameters ($\hat{R}_{\max} = 1.002$). The effective sample size ratio $N_{\text{eff}}/N_{MC}$ is 1.0 for all parameters (c.f., 0). The QP is solved on the commercial solver CPLEX.

The optimisation solver and the MCMC sampler (4 chains with 2500 draws each) complete in 0.1s and 60s, respectively, on an Intel 2.9 GHz i5 CPU, 16 GB DDR3 RAM laptop computer. Therefore, the optimization approach outperforms the Bayesian approach when applied to very small transit systems like test network A, if estimates' uncertainty is not required. In contrast, the Bayesian approach provides a measure of the estimates' uncertainty and considers the concentration of probability mass of the parameter posterior distribution, which reveals essential in high dimensional problems. In large-scale transit networks, we therefore focus on the Bayesian approach. Moreover, we prompt to Section 4.3, where we revisit the QP approach on a high-dimensional parameter estimation problem for a large-scale network and find that the QP approach produces unreliable OD-estimates, regarding the coherence with the model assumptions.

### 4.1.2 Effect of count data dispersion on accuracy and precision

The dispersion parameter $\phi$ controls the variance of the entry count observations. We vary the value of $\phi$ considering $N = 30$ observations and constant entry count means given in Appendix E, Table E.1. Fig. 6 demonstrates that an increase in the variance of the entry count observations (i.e., the distribution of entry counts becomes more overdispersed) improves the accuracy of the Bayesian and QP estimates and the precision of the Bayesian estimates.

The OD-coefficient estimates are solely distilled from the entry and exit count observations and rely on identifying how changes in the entry counts at one station influence the changes in the exit counts at another station. Therefore, if the $N$ observations of entry counts at all stations were constant, there would be infinitely many solutions to the OD-estimation problem. Conversely, as the variability of the entry counts becomes more pronounced, the signals that entry count fluctuations impose on the exit count observations are more discernible and enable the OD-estimation method to more accurately and precisely infer the OD-coefficients.

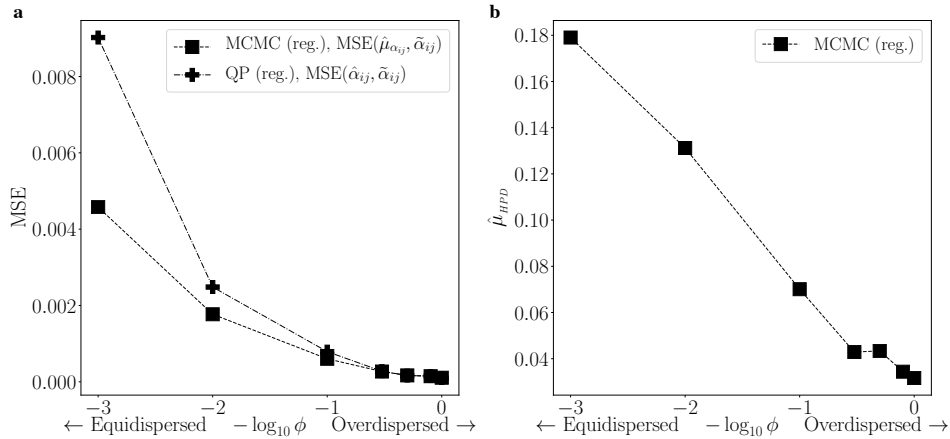

**Figure 6.** The influence of count dispersion on the OD-estimates' accuracy and precision. (a) The mean squared error with respect to the OD-coefficients' true values, given the Bayesian posterior mean and the QP point estimate, for different dispersion levels of the station entry count observations; (b) The mean of the Highest Posterior Density (HPD) intervals across all posterior coefficient estimates for different dispersion levels of the station entry count observations. Estimates are based on the regularised (reg.) Bayesian and QP model formulations.

### 4.2 Instantaneous-balance and average-delay model sensitivity

We apply the average-delay model and the instantaneous-balance model to test network B and compare their estimation results. For test network B, the cumulative entry and exit counts may be imbalanced over the observation window. To comply with the instantaneous-balance assumption, we correct for this imbalance using the approach in Appendix D prior to applying the instantaneous-balance model. For this discussion, we solely focus on the Bayesian estimates. We analyze the sensitivity of the OD-estimation accuracy and precision with respect to (a) the model type, i.e. the instantaneous-balance model and the average-delay model, (b) the observation window width $w$, (c) the trend scale $\eta$, and (d) the count data dispersion $\phi$. Fig. 7 shows the accuracy and precision of the OD-coefficient estimates for the instantaneous-balance and the average-delay model; markers close to the origin indicate accurate and precise estimates.

Fig. 7a shows that the average-delay model estimates are overall more accurate and precise. Therefore, the average travel delay model is more reliable in retrieving the true OD-matrix from the time-binned count data than the time-aggregated instantaneous-balance model and is more robust to changes in count characteristics. Additionally, Fig.'s 7b to 7d illustrate the effect of observation window width, trend scale, and Negative Binomial dispersion, respectively. The arrows in Fig. 7 indicate the stepped average change of the accuracy and precision given a change of the respective test parameter values. The observation window width significantly influences the accuracy and precision of both the instantaneous-balance and average-delay model estimates, since it prescribes the information horizon that the models are able to observe and the number of binned data points considered in the average-delay model. The trend scale unnoticeably influences the results of the average-delay model, while the instantaneous-balance struggles to determine reliable parameter estimates as the trend scale increases and interferes with the instantaneous-balance assumptions. Moreover, the estimates of the models improve as the count dispersion increases, because stronger fluctuations in the count observations provide clearer evidence of the level of coupling between origin and destinations as shown in Section 4.1.2. Appendix I further supports these findings based on the correlation and metamodel-based sensitivity analysis between the test parameters and the accuracy and precision statistics.

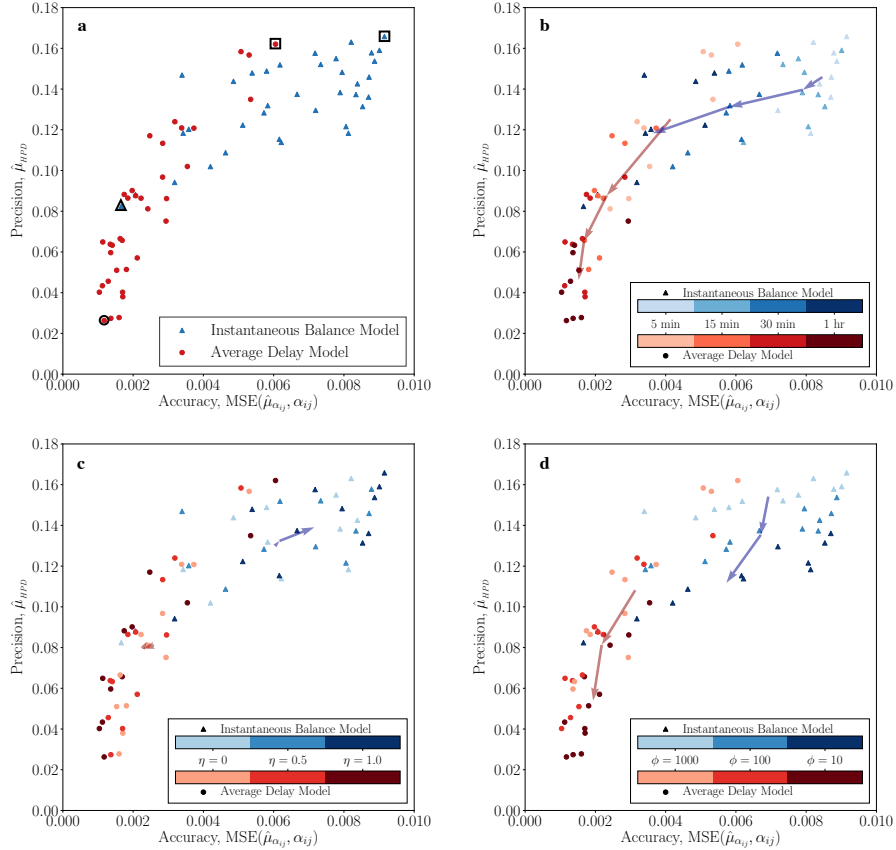

**Figure 7.** The effect of (a) model type (instantaneous-balance or average-delay), (b) observation window width $w \in \{5, 15, 30, 60\}$ min, (c) trend scale $\eta \in \{0.0, 0.5, 1.0\}$, and (d) count dispersion $\phi \in \{10, 100, 1000\}$ on the accuracy and precision of the Bayesian OD-estimates with $N = 100$ observations, resulting in $2 \cdot 4 \cdot 3 \cdot 3 = 72$ test sites. The arrows indicate the average direction of the change in accuracy and precision of the two model types as (a) the observation window width increases, (b) the trend scale increases, and (c) the count dispersion increases (i.e., as $\phi$ decreases). The boxed markers in (a) indicate the worst estimate of the instantaneous-balance and average-delay model. The triangular-framed and circled markers in (a) indicate the best estimate of the instantaneous-balance model and of the average-delay model, respectively; these are further discussed in 0.

The estimation ability of the average-delay model, however, demands specific data requirements and is subject to higher computational complexity. In particular, the average-delay model relies on granular time series and extensive information regarding the transit network timetable and transfer times between stations. Furthermore, while both the average-delay and instantaneous-balance model require matrix evaluations associated with a time complexity of $\mathcal{O}(n^{2.373})$ to $\mathcal{O}(n^3)$ depending on the specific algorithm, the average-delay model requires $S$ looped matrix evaluations for the subsets of the OD-assignment matrix for each destination station. For 100 observations on test network B, the execution time of the MCMC sampler varies between 1 minute to 30 minutes for the average-delay model and between 30 seconds to 3 minutes for the instantaneous-balance model on an Intel 2.9GHz i5 CPU, 16 GB DDR3 RAM laptop computer. The variability of the execution time is subject to the combinations of different observation window widths, trend scales, and Negative Binomial dispersions, creating irregularly complex posterior distributions to sample from.

In case the computations for the average-delay model are prohibitively costly (e.g., in the case of the NYC subway network), or travel time information is not readily available, the instantaneous-balance model shows to perform well for (a) a sufficiently long observation window to span the travel times between origins and destinations, (b) flat or small trends in the periodically repeating (e.g., daily) count series, (c) a suitable degree of dispersion in the count series and aggregated observed data, and (d) an ample number of observations across which system conditions (travel demand and supply) can be assumed to be invariant.

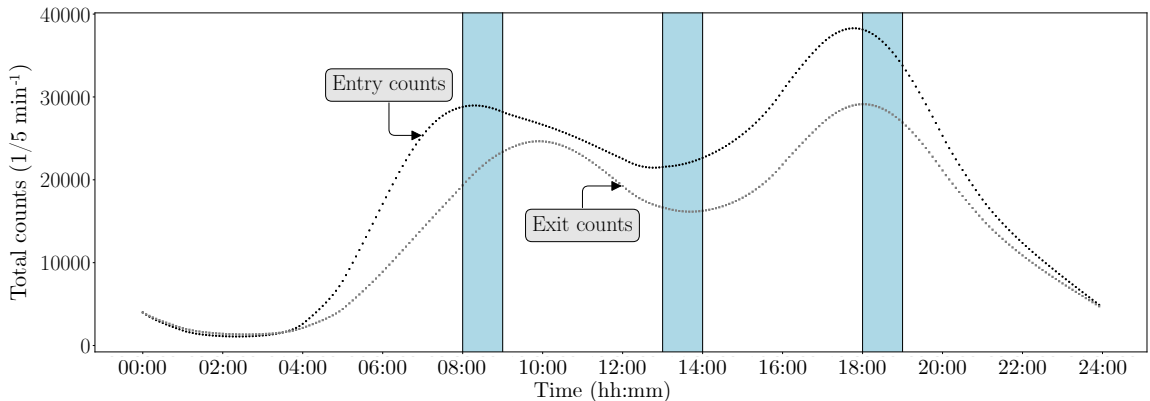

**Figure 8.** The recorded average total number of passengers entering and exiting across all stations of the NYC subway network. The blue vertical bars indicate the morning peak, midday off-peak, and evening peak observation window.

### 4.3 OD-estimation for the NYC subway network

We apply the proposed instantaneous-balance Bayesian OD-estimation model to the NYC subway network. With $S = 471$ stations, the total number of OD-coefficient estimates is equal to $S(S - 1) = 221370$. We consider all weekday counts in three observation windows, namely, 8 AM to 9 AM, 1 PM to 2 PM, and 6 PM to 7 PM between April and October with records starting in 2010 and ending in 2018, resulting in $N = 1315$ observations. The three windows characterize the morning peak, the midday off-peak, and the evening peak, respectively, and are shown in Fig. 8 along with the average total number of entering and exiting passengers throughout the day. According to the timetable data in (Metropolitan Transport Authority, 2019b), the average travel time across the network is roughly 55 minutes and therefore justifies the width of the one hour observation window for the morning peak, midday off-peak, and evening peak window. Passenger count data for NYC are only available in an aggregated 4-hour resolution basis and, therefore, are solely applicable to the instantaneous-balance OD-estimation model. Moreover, in preliminary testing it was found that using approximately ten days of count observations together with the average-delay formulation results in approximately the same computation time as using the full 1315 daily observations with the instantaneous-balance formulation.

The posterior distribution estimates of the OD-coefficients are determined from the MCMC sampling algorithm described in Section 0. The MCMC sampler completes in roughly 4 days for 2000 iterations (1000 warm-up and 1000 sampling iterations) on each of 10 parallel chains, when executed on a computing cluster with 64 Intel® Xeon® IE5-2699 v3 2.30GHz CPU's and 200 GB of DDR3 RAM. This results in a sampling with a total of 10000 MCMC draws from the (marginal) posterior distribution of each of the 221370 OD-split coefficients for each observation window. Regarding the NUTS sampler, no divergences are recorded during the sampling runs of all three observation windows. The tree search depth of the NUTS algorithm is set to 10. The sampler saturates this tree depth for all draws in all three sampling runs. While the absence of divergences indicates that the model is well specified and the associated MCMC draws are valid, tree-depth saturations indicate that the algorithm is not sampling efficiently. Nonetheless, the drawn samples are valid MCMC draws, even if the search tree of the NUTS algorithm saturates at the maximum defined depth.

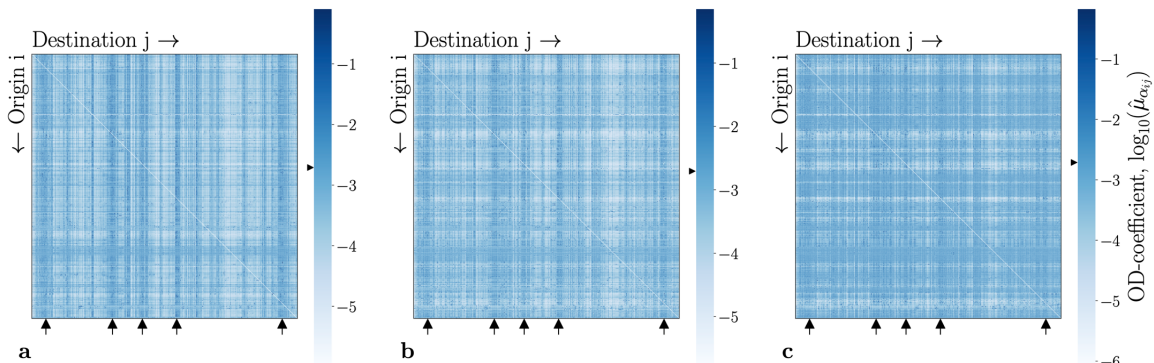

**Figure 9.** OD-matrices of the sample means of the OD-coefficient posterior for the (a) morning peak, (b) midday off-peak, and (c) evening peak window. The markers left of the colour-scales indicate the OD-coefficient prior mean equal to $(S - 1)^{-1}$. Vertical arrows below each plot correspond to the neighbourhood of stations that consistently observe more than 3000 exiting passengers per hour. Rows and columns are sorted according to the geographic locations of stations and the stop patterns of services running through them.

### 4.3.1 OD-coefficient posterior means and absolute OD-demand

Fig. 9 visualizes the matrices of the OD-coefficient posterior means for the NYC subway network and the morning, midday, and evening window. Fig. K.1 in Appendix K additionally maps these OD-coefficient means onto the physical locations of the NYC subway network. Rows and columns of the OD-matrix correspond to the 471 stations in the NYC subway network, sorted and grouped geographically and by the subway lines that pass through them. The OD-coefficient posterior means vary distinctly across origins and destinations, with noticeable patterns of smaller and larger valued estimates. In comparison, the Dirichlet OD-coefficient hyperprior in Eq. (8) defines a mean equal to $(S-1)^{-1} \approx 0.002$ for all coefficients. Since the concentration parameter in Eq. (8) is set to $c = 1$, the simplex-constrained OD-coefficient support is subject to uniform probability mass throughout. The distinct OD-coefficient patterns therefore indicate that the OD-coefficient posterior sampling trajectory is mostly guided by the likelihood model and not the hyperprior. Moreover, the maximum OD-coefficient posterior means are approximately 0.7 in all observation windows, indicating that consistently certain stations attract a dominant proportion of passenger trips from other stations.

The three OD-matrices in Fig. 9 display vertical bands of OD-coefficients with mean estimates predominantly in the range between 0.008 to 0.012. These bands are marked by the vertical arrows at the bottom of each plot and correspond to the destination stations that consistently observe more than 3000 exiting passengers per hour across all three observation windows and, thus, persistently attract passenger trips throughout the day[3]. However, the patterns of dominant bands are visually more dissolved in Fig. 9c as compared to Fig. 9a. This is because a large part of the morning peak demand reverses while only some of the demand continues to be bound for stations that observe large demand throughout the day.

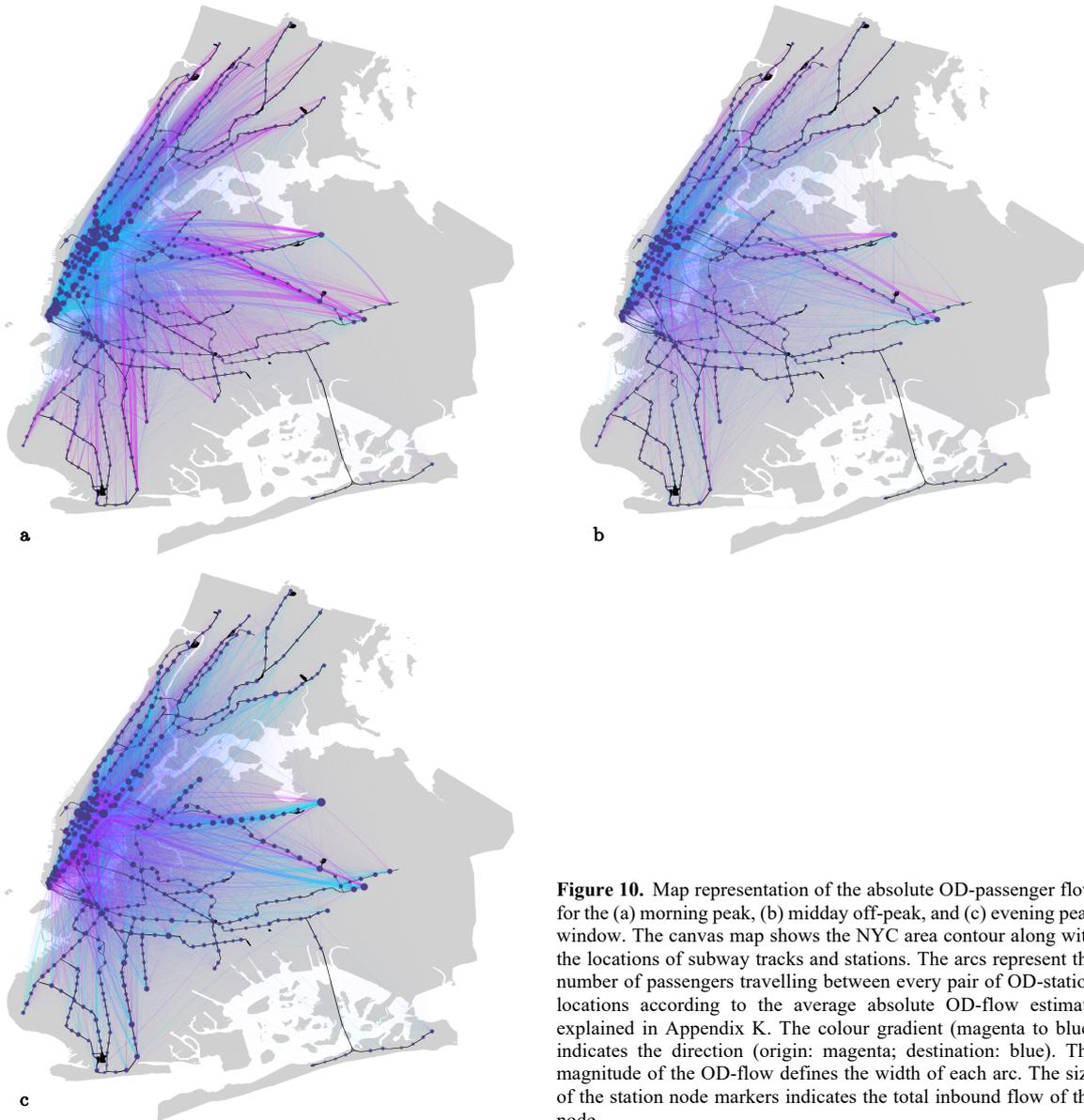

**Figure 10.** Map representation of the absolute OD-passenger flow for the (a) morning peak, (b) midday off-peak, and (c) evening peak window. The canvas map shows the NYC area contour along with the locations of subway tracks and stations. The arcs represent the number of passengers travelling between every pair of OD-station locations according to the average absolute OD-flow estimate explained in Appendix K. The colour gradient (magenta to blue) indicates the direction (origin: magenta; destination: blue). The magnitude of the OD-flow defines the width of each arc. The size of the station node markers indicates the total inbound flow of the node.

---

[3] These stations are: Grand Central - 42 St (lines 4,5, and 6), 34 St - Herald Sq (lines B and D), 34 St – Penn Station (lines A, C, and E), 34 St – Penn Station (lines 1, 2, and 3), and 14 St - Union Square (lines N, Q, R, and W)

Indeed, this becomes more evident in the absolute OD-flow. Whereas Fig.'s 9 and K.1 plot the OD-coefficient posterior means that quantify the relative passenger trip productions of each origin station, Fig. 10 shows the absolute passenger flow between origin and destination locations following the estimate of the average absolute

OD-flow detailed in Appendix K. Fig. 10a demonstrates that during the morning peak, most passenger trips are bound for stations along the subway line corridors that stretch across the business and commercial centres of Upper East and West Side, Midtown, and Lower Manhattan. During the midday off-peak in Fig. 10b, the passenger ridership reduces, while most passenger trips continue to be bound towards stations in Manhattan from stations in the surrounding boroughs. Moreover, stations in Queens and the Bronx start to more strongly attract passenger trips. During the evening peak in Fig. 10c, the number of passenger trips bound towards stations in Lower and Midtown Manhattan reduces, while a large number of passenger trips end at stations located in residential areas in Brooklyn, Queens, and the Bronx. Only few stations in Lower and Midtown Manhattan continue to attract many passenger trips, due to their vicinity to regional train services and activity centers.

### 4.3.2 Posterior predictive checks

The instantaneous-balance model predictions and the real-world observations for the NYC subway are compared using posterior predictive checks, in order to establish whether the model can reproduce the original observations. To this aim, the means of the observed exit counts $\bar{y}_j$ are compared to predictions generated from the instantaneous-balance model formulation in Eq. (6). The intercept $r_j$ is omitted, so that the predictive model satisfies the uncorrected instantaneous-balance assumption. In this way, it is assumed that passenger exit counts are solely the result of passengers entering the network and distributing across the destinations according to the relative proportions given by the OD-coefficients. Moreover, the predictive model determines the predictive distribution of the uncorrected latent location parameter $\hat{\mu}_{y,j}$, as opposed to the predictive distribution of the exit count observations. The average entry counts of the 1315 observations are substituted, as well as a thinned sample of 1000 draws from the OD-coefficient posterior.

Fig. 11 compares the mean of the observed exit counts $\bar{y}_j$ with the posterior predictive distributions of the exit count location parameter $\hat{\mu}_{y,j}$ (for large exit counts this converges to the mean). The coefficient of variation is less than 0.04 across all predictive distributions. Consequently, the predictive distribution samples (i.e., 1000 samples for each station) that are plotted in Fig. 11 are narrowly bundled for every station, visually almost converging to a single point. The model closely captures the observed data across all three observation windows, particularly for stations with large exit counts. Indeed, for 382, 419, and 452 of the 388, 421, 452 stations with mean exit counts of ≥ 100 passengers per hour for the morning, midday, and evening window, respectively, the relative absolute prediction errors are below 10%. However, one particular station exhibits anomalous prediction errors; for Wall Street station (lines 4 and 5) the relative prediction errors are approximately equal to 25%, despite generally large exit count observations, i.e. the observed mean exit counts during the morning, midday, and evening observation window are in the order of 9500, 2500, and 1500 passengers per hour. We attribute these discrepancies to inconsistent data records of the entry and exit count observations at Wall Street station, where in 2014 the observations significantly dropped (c.f., Fig. G.1d in Appendix G), and thus violate the assumptions regarding the periodicity of recurring system conditions. Finally, at stations with exit counts roughly below 100 passengers per hour, the model over-predicts by up to almost four-fold (i.e., 300% relative error). These errors are negligible, considering that the total exit counts sum to approximately $3 \cdot 10^5$ to $5 \cdot 10^5$ passengers per hour.

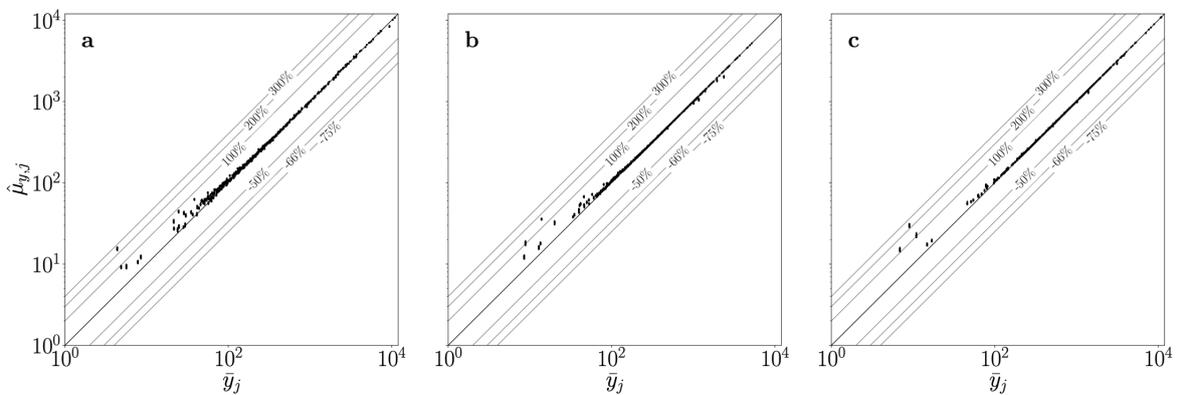

**Figure 11.** Instantaneous-balance model posterior predictive exit count distributions versus the observed exit counts for the 471 stations of the NYC subway network during the (a) morning peak, (b) midday off-peak, and (c) evening peak window. For each station the distribution consists of 1000 samples. The grey annotated lines indicate the (relative) percentage difference contours.

### 4.3.3 MCMC convergence and autocorrelation

Convergence monitoring is essential in qualifying whether the MCMC samples are drawn from the equilibrium distribution, i.e., each randomly-initialised chain has converged to the equilibrium distribution. This is quantified by the potential scale reduction statistic, $\hat{R}$, as the ratio between the sum of the average within-chain variance plus cross-chain sample variances, and the average within-chain variance (Gelman et al., 1992). A value of $\hat{R}$ close to unity indicates that the chains of the MCMC algorithm sample from the same equilibrium posterior distribution. Fig. 12 shows the distribution of $\hat{R}$ for the 221370 OD-coefficients and the three observation windows. Fig.'s 12a and 12b show that most $\hat{R}$ statistics are smaller than 1.2 and predominantly concentrate close to 1.0 for the morning and midday observation window. Hence, the sampling chains converge for 99.96% and 99.89% of the 221370 OD-coefficients for the morning and midday observation windows, respectively, considering $\hat{R} = 1.2$ as the convergence threshold. However, Fig. 12c shows that the evening observation window only exhibits convergence for few OD-coefficients for the same $\hat{R}$ threshold (constituting 5.18% of all coefficient posteriors), where the non-converged samples account for approximately 98% of the overall OD-flow between stations. Therefore, the posterior distribution for the evening observation window is more difficult to explore and sample from. The count observations show that data records are slightly less dispersed for the evening observation window than for the morning and midday observation windows, which could be the cause for the sampling deficiencies. Despite this, the results for the evening observation window are considered sufficiently reliable, because the posterior predictive check in section 4.3.2 indicates that the model adequately captures the observations for the evening observation window.

Finally, autocorrelation analysis shows that 95% of all coefficient estimates exhibit an effective sample size ratio $N_{\text{eff}}/N_{MC} > 0.85$ across the three observation windows and 80% of the estimates are associated with $N_{\text{eff}}/N_{MC} > 0.95$. Therefore, the estimates' variance can mostly be attributed to the inherent observation variability and modelling uncertainties rather than to autocorrelated samples. 0 further elaborates on the effective sample size.

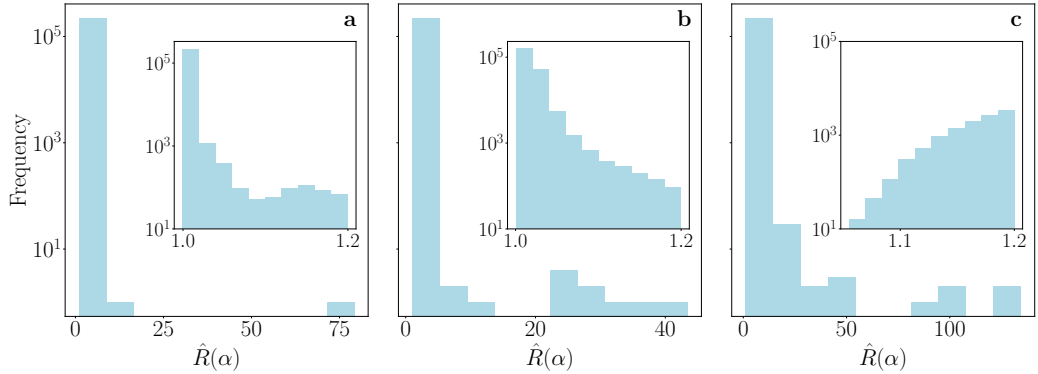

**Figure 12.** The distribution of the $\hat{R}$ potential scale reduction statistic of the OD-coefficient posterior for the (a) morning peak, (b) midday off-peak, and (c) evening peak window for the NYC subway network. The insets show the distributions for estimates with $\hat{R} \leq 1.2$. $\hat{R}$ values of OD-coefficient posteriors with a mean smaller than $10^{-9}$ are not included in the analysis.

### 4.3.4 Inadequacy of QP estimates

Fig. 13 illustrates the OD-coefficient matrix resulting from the QP optimization model in Appendix A and applied to the NYC subway system with observations from the morning observation window. The QP model is solved in CPLEX resulting in an optimal solution after approximately 6 hours on the same computing cluster as for the MCMC sampler. Despite the generally favourable non-sparse output of QP optimization models, the QP optimization produces a highly sparse OD-matrix, with a matrix sparsity of 99% (for a zero round-off for values below $10^{-9}$). In contrast, the MCMC-estimates in Fig. 9a result in a matrix sparsity of 0.23%, where 64% of the coefficients are larger than $10^{-3}$. Moreover, the maximum coefficient for the QP-derived OD-matrix is equal to 1.0, suggesting that all inflow at an origin station is assigned to a single destination station. However, the passenger inflow at an origin station is likely to distribute to multiple destination stations – even those that do not dominantly attract many passenger trips. Most of the OD-flow according to the QP derived OD-matrix is restricted to only a few OD-pairs, while the MCMC-derived matrix distributes origin inflow considerably more evenly. A possible cause of this discrepancy can be that QP optimization is sensitive to outliers in the observed data. However, outliers were rectified prior to solving the optimization problem (see Appendix G) and tests with a linear programme formulation, that is less sensitive to outliers, similarly result in a highly sparse result. Another cause is that the high-dimensional problem, combined with the limited number of count measurements (i.e., the underdetermined problem of 1315 observations versus $471 \cdot 471 = 221841$ OD-coefficients), may have multiple local optima, such that the OD-matrix estimate shown in Fig. 13 is only one of numerous possible solutions. Moreover, the high-dimensional underdetermined problem, which leads to overfitting of the observed data and difficulty in retrieving a possible global optimum, is additionally hampered by model assumptions, that may partly mis-specify the real-world processes. For this reason, the QP optimization routine presumably fails to adequately determine physically interpretable estimates of the OD-coefficients in this specific problem.

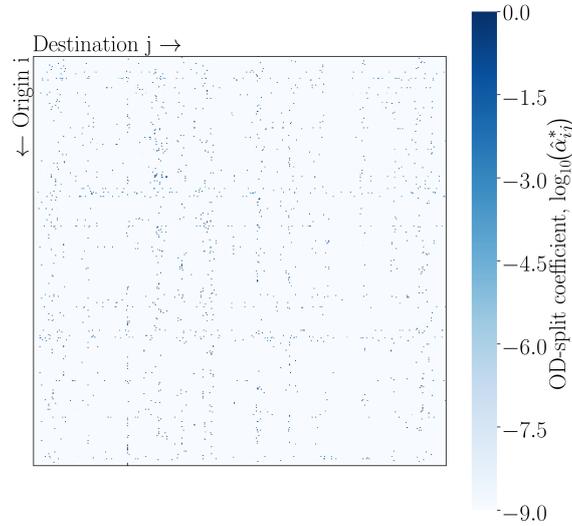

**Figure 13.** The inferred OD-matrix for the NYC subway system for the morning peak window with $N = 1315$ observations, based on the quadratic programme optimisation approach. OD-coefficient estimates with $\hat{\alpha}_{ij}^* < 10^{-9}$ are plotted as $10^{-9}$.

## 5. Conclusion

This work proposes a Bayesian inference approach for static Origin-Destination (OD)-estimation and system identification from time series measurements of nodal in- and outflows in a networked transit system. The approach is particularly suited for high-dimensional and underdetermined problems with several hundred network nodes.

We suggest two different model formulations: the instantaneous-balance and average-delay model. Overall, the average travel delay model is more robust to changes in the observation characteristics and is able to more accurately and precisely determine the OD-coefficient posterior estimates, under the condition that high-resolution count data are available. However, given the appropriate observation characteristics, the instantaneous-balance model recovers the OD-matrix sufficiently well at smaller computational cost.

Moreover, we reformulate and extend the instantaneous-balance model into a constrained quadratic programming (QP) optimization problem for benchmarking the Bayesian approach. We show that the estimation accuracy of the Bayesian posterior means is comparable to that of the QP point estimates on a contrived test network. However, the QP point estimates provide only the residual information between the predictions and observations, as opposed to the Bayesian approach, which considers the concentration of mass of the parameter posterior distribution. This is essential in higher dimensional problems that often suffer from the curse of dimensionality. Moreover, the Bayesian posterior estimates provide a measure of the parameter uncertainty, which translate into a generalisable assessment of the prediction quality of the model.

At last, we test the Bayesian and QP approach on the NYC subway system. The QP approach succumbs to the high-dimensionality of this large-scale real-world test network, and thus fails to provide real-world interpretable OD-coefficient estimates that comply with engineering intuition. Further research will be required to establish the sensitivity of the QP formulation to the problem size and properties of the observational data. Especially, we do not investigate under which specific conditions the QP approach ceases to provide accurate estimates – In this regard, this work merely identifies an example for which the QP formulation is unable to provide dependable estimates. In contrast, the Bayesian instantaneous-balance model results in OD-coefficient estimates for the NYC subway system that overall reflect intuitive demand patterns alongside satisfying convergence, autocorrelation, and posterior predictive tests. Given the incompleteness and coarseness of the turnstile records and the time complexity constraints, the average-delay formulation is ruled out as a possible model for the real-world test scenario.

With the advent of novel fare collection systems, transit OD-estimation is becoming considerably more refined and the presented work will be complementary to smart-card-enabled methods that can extract OD-information from trip records. Moreover, we propose that this work is not exclusive to transit network OD-estimation, and is presumably equally amenable to road traffic OD-estimation or network tomography.

**Acknowledgements**
-

**Funding**
This work was supported by the National Research Foundation (NRF) of Singapore [FI 370074011], as part of NRF's Campus for Research Excellence and Technological Enterprise (CREATE) program.

**Declaration of conflict of interest**
The authors declare no conflict of interest.

**Glossary**

| | |
|---|---|
| ARIMA | Auto-Regressive Integrated Moving Average |
| CPU | Central Processing Unit |
| DDR3 | Double Data Rate Type 3 |
| GB | Giga Byte |
| GTFS | General Transit Feed Specification |
| HMC | Hamiltonian Monte Carlo |
| HPD | Highest Posterior Density |
| INGARCH | Integer-valued Generalised Autoregressive Conditional Heteroscedasticity |
| MCMC | Markov Chain Monte Carlo |
| MNL | Multinomial Logit |
| MSE | Mean Squared Error |
| MTA | Metropolitan Transport Authority |
| NYC | New York City |
| NUTS | No U-Turn Sampler |
| OD | Origin-Destination |
| QP | Quadratic Programme |
| RAM | Random Access Memory |

**Notation**

| | |
|---|---|
| $\boldsymbol{A}$ | OD-matrix |
| $B$ | Batch size of autoencoder neural network |
| $C$ | Normalizing constant |
| $\mathcal{D}$ | Observational evidence |
| $F_t$ | INGARCH model trend process |
| $\boldsymbol{H}$ | OD-assignment matrix of size $K \times S$ |
| $\boldsymbol{H}_k$ | Sub-matrix of $\boldsymbol{H}$ of size $N \times T_a \times S$ |
| $H_t$ | Non-negative ARIMA process |
| $K$ | Total number of rows of the OD-assignment matrix equal to $S \times N \times T_a$ |
| $M$ | Sample size of contiguous records for autoencoder training (i.e., the sub-frame width) |
| $N$ | Number of observations |
| $N_{MC}$ | Number of (Markov chain) Monte Carlo samples |
| $N_{nn}$ | Number of records within one batch of a autoencoder neural network training sample |
| $N^+(\mu, \sigma)$ | Lower zero-truncated normal distribution, given location parameter $\mu$ and scale parameter $\sigma$ |
| $\hat{R}$ | Potential scale reduction statistic |
| $S$ | Total number of stations |
| $T$ | Total number of time bins per observation window |
| $T_a$ | Total number of arrival time bins per arrival window |
| $V_k$ | Deterministic utility component of path alternative $k$ |
| $\boldsymbol{X}$ | Entry count matrix with size $N \times S$ |
| $\boldsymbol{Y}$ | Exit count matrix with size $N \times S$ |
| $\boldsymbol{Z}_t$ | INGARCH model covariate vector |
| | |
| $c$ | Symmetric Dirichlet concentration parameter |
| $e_j$ | Residual between the predicted and the observed exit count mean at station $j$ |
| $h_{ki}$ | Element of the OD-assignment matrix $\boldsymbol{H}$ at index $(k, i)$ |
| $i$ | Origin station index |
| $j$ | Destination station index |
| $n$ | Observation index |
| $p(\bullet\|\bullet)$ | Likelihood probability density |
| $r_j$ | Intercept parameter for station $j$ |
| $t_a$ | Arrival time bin index |
| $t_d$ | Departure time bin index |
| $t_0$ | Start of the departure window |
| $t_1$ | Start of the arrival window |
| $\Delta \bar{t}_{ij}$ | Average travel delay (in number of time bins) between origin $i$ and destination $j$ |
| $q$ | Placeholder variable |
| $w$ | Observation window width |
| $\bar{x}_i$ | Observed mean entry counts at station $i$ |
| $x_i^{(n)}$ | Observed entry counts at station $i$ during observation instance $n$ |
| $x_i^{(n,t_d)}$ | Observed entry counts at station $i$ during observation instance $n$ and departure time bin $t_d$ |
| $x_t^{(n)}$ | Total observed entry counts across all stations during observation $n$ |
| $y_t^{(n)}$ | Total observed exit counts across all stations during observation $n$ |
| $x_+^{(n)}$ | Excess entry counts during observation $n$ |
| $\tilde{x}_i^{(n)}$ | Corrected observed entry count at station $i$ during observation $n$ |
| $\boldsymbol{y}_j$ | Column vector that contains the number of exiting passengers at destination $j$ across all time bins and observations; The size is $N \times T_a$ |
| $\bar{y}_j$ | Mean exit count observations at station $j$ |
| $y_j^{(n)}$ | Observed exit count at station $j$ during observation instance $n$ |
| $y_j^{(n,t_a)}$ | Observed exit count at station $j$ during observation instance $n$ and arrival time bin $t_a$ |
| $\tilde{y}_j^{(n)}$ | Corrected observed exit count at station $j$ during observation $n$ |
| $\boldsymbol{z}_k$ | Utility vector of path alternative $k$ |
| $\alpha_{ij}$ | OD-coefficient between station $i$ and station $j$ |
| $\tilde{\alpha}_{ij}$ | True OD-coefficient between station $i$ and station $j$ |
| $\hat{\alpha}_{ij}^*$ | Point estimate of the OD-coefficient $\alpha_{ij}$ |
| $\boldsymbol{\alpha}_{i\bullet}$ | Simplex-constrained row vector corresponding to the $i$-th row of the OD-matrix $\boldsymbol{A}$ |
| $\tilde{\boldsymbol{\alpha}}_{i\bullet}$ | Simplex-constrained row vector of the true OD-coefficient matrix for station $i$ |
| $\boldsymbol{\alpha}_j$ | Column vector corresponding to the $j$-th column of the OD-matrix $\boldsymbol{A}$ |
| $\beta_k$ | Regression parameter |
| $\gamma$ | INGARCH model covariate weight parameters |
| $\delta_0^{t-\tau}(t=\tau)$ | INGARCH model deterministic intervention covariate |
| $\epsilon$ | Error |
| $\varepsilon_j^{(n)}$ | Residual between the predicted and the observed exit count at station $j$ during observation $n$ |
| $\zeta$ | MNL model attribute coefficient (weight) vector |
| $\eta$ | Trend scale parameter |
| $\theta$ | Bayesian model parameter |
| $\lambda_t^{(n)}$ | INGARCH model mean covariate at time $t$ during observation $n$ |
| $\mu$ | Mean |
| $\mu_{y,j}^{(n)}$ | Location parameter of exit count likelihood model at station $j$ during observation $n$ |
| $\hat{\mu}_{\alpha_{ij}}$ | Estimate of the OD-coefficient $\alpha_{ij}$ posterior distribution mean |
| $\mu_{x,i}$ | Mean of Negative Binomial observational entry count model for station $i$ |
| $\hat{\mu}_{y,j}$ | Uncorrected latent location parameter of the truncated normal distribution exit count likelihood model for station $j$ |
| $\nu_0$ | INGARCH model intercept |
| $\nu_k$ | INGARCH model observation regression parameter |
| $\xi_l$ | INGARCH model mean regression parameter |
| $\pi(\bullet\|\bullet)$ | Posterior probability density |
| $\pi(\bullet)$ | Prior probability density |
| $\sigma_{y,j}$ | Observation-invariant scale parameter of exit count likelihood model at station $j$ |
| $\sigma^2$ | Variance |
| $\tau$ | Time occurrence of INGARCH model intervention |
| $\phi$ | Negative Binomial dispersion parameter |
| $\omega_0$ | INGARCH model intervention scaling parameter |

# Appendices

## Appendix A     The instantaneous-balance optimization model

The proposed Bayesian inference model is compared with an equivalent point estimate formulation. The point estimate formulation finds OD-coefficients estimates, such that the total sum of the squared residuals between the model predictions and observations is minimised. It is formulated as a constrained quadratic optimization programme that minimizes the $L_2$-norm (i.e., sum-squared-error objective function) between the predicted and observed exit counts, where the formal expressions are given by

$$\underset{A}{\text{minimize}} \sum_{\substack{j \in S \\ n \in N}} \left(\varepsilon_j^{(n)}\right)^2 + \sum_{j=1}^{S}\left(r_j\right)^2 + \sum_{j=1}^{S}\left(e_j\right)^2$$

$$\text{subject to} \quad \sum_{\substack{i=1 \\ i \neq j}}^{S} \alpha_{ij} x_i^{(n)} + r_j - y_j^{(n)} = \varepsilon_j^{(n)} \quad \forall \quad j \in \{1, 2, \dots, S\}, \quad n \in \{1, 2, \dots, N\},$$

$$\sum_{\substack{i=1 \\ i \neq j}}^{S} \alpha_{ij} \bar{x}_i - \bar{y}_j = e_j \quad \forall \quad j \in \{1, 2, \dots, S\}, \quad [\text{A.1}]$$

$$\sum_{\substack{j=1 \\ j \neq i}}^{S} \alpha_{ij} = 1 \quad \forall \quad i \in \{1, 2, \dots, S\},$$

$$0 \leq \alpha_{ij} \leq 1 \quad \forall \quad i \neq j.$$

Eq. (A.1) is similar to the constrained optimization model by Cremer and Keller (1987), who used if for OD-estimation of the turning movements at intersections. We extend the formulation by adding the intercept $r_j$ and including the minimal-bias and expected value regularisation terms (i.e., $\sum_{j=1}^{S}(r_j)^2$ and $\sum_{j=1}^{S}(e_j)^2$) into the objective function. These terms are further explained in Section 2.2.2, which also describes their implementation into the Bayesian model formulation.

    The optimisation objective function could similarly be expressed in terms of the sum-absolute-error (i.e., the $L_1$-norm), resulting in a linear programme formulation. The linear and quadratic formulation each come with canonical properties that make them favourable depending on the problem setting and desired outcome. The least absolute deviations problem associated with the linear programme is robust to outliers in the data and produces sparse outputs, whereas the least squares problem associated with the quadratic programme is sensitive to outliers and tends to produce non-sparse outputs. While it is reasonable to assume that the inflow at an origin will be dominantly bound for only a few destinations, it is implausible that none of the inflow will ever be bound for the remaining destination nodes. Hence, a sparse solution that results in many zero-valued OD-coefficients is neither desired nor plausible. We therefore choose the quadratic programme formulation in Eq. (A.1) and rectify data outliers prior to solving it.

## Appendix B     The Dirichlet hyperprior

Section 2.2.1 describes that we place a Dirichlet hyperprior on the rows of the OD-matrix. Here, we illustrate the influence of the concentration parameter $c$ included in the hyperprior. Fig. B.1 shows samples from a Dirichlet distribution according to $\text{Dir}(x; c[1,1,1])$. This particular distribution spans a two-dimensional simplex equivalent to the triangle with vertices $[1, 0, 0]$, $[0, 1, 0]$, and $[0, 0, 1]$. By reducing the concentration parameter $c$ below 1, the variates are biased to be sparser, whereas $c > 1$ creates denser, hence concentrated, variates. For the MCMC-based OD-estimation problem, values of $c$ considerably below 1 may make the simplex prohibitively difficult to sample over, given the concentration of probability mass near the vertices of the simplex. A value larger than 1, may put too much bias towards equal OD-coefficients.

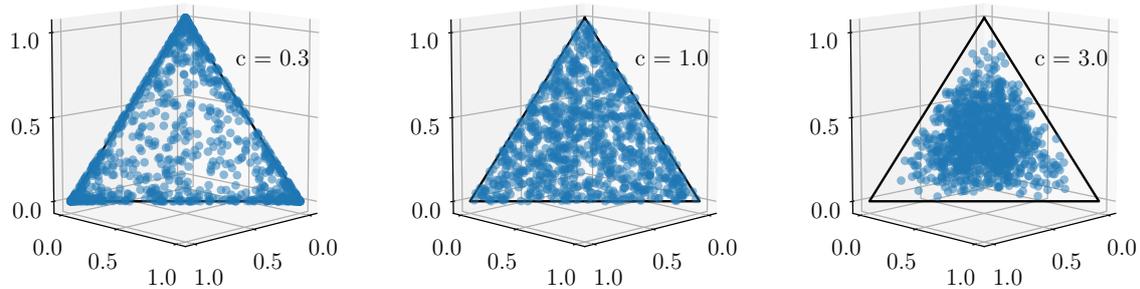

**Figure B.1.** Samples from a Dirichlet distribution with $\text{Dir}(x; c[1,1,1])$. Variates are either more or less concentrated towards the vertices of the $(k-1)$-dimensional simplex (with $k = 3$), depending on the value of the concentration parameter $c$. For $c = 1$ the samples are uniformly distributed over the simplex.

**Appendix C      Formal definition of expectation**

The expectation is determined by integrating the product of the parameter-dependent function $f(\theta)$ with the parameter posterior probability density $\pi(\theta|\mathcal{D})$ over all possible parameter values $\theta$, according to

$$\mathbb{E}[f(\theta)] = \int_{\Theta} f(\theta)\pi(\theta|\mathcal{D})d\theta \,, \qquad [C.1]$$

where $\mathcal{D}$ are evidential observations or measurable data, and the term $\pi(\theta|\mathcal{D})$ is the unknown posterior distribution. Since we are unable to determine a closed-form solution for the posterior distribution, we resort to the estimator for the expectation, given by

$$\mathbb{E}[f(\theta)] \approx \frac{1}{N_{MC}} \sum_{n=1}^{N_{MC}} f\big(\theta^{(n)}\big) \,, \qquad [C.2]$$

where $\theta^{(n)}$ are draws from the posterior distribution, which are determined from the MCMC sampling routine.

**Appendix D      Count imbalance correction**

As a consequence of the cumulative count aggregation for the instantaneous-balance model, the entry count observations are not guaranteed to balance with the exit count observations that are accumulated over the same observation window. A possible solution to re-balance these counts is to distribute the excess total counts across the stations by rule of proportion, accounting for the stations' relative contribution to the total entry and exit counts, such that

$$\begin{cases} \tilde{x}_i^{(n)} = x_i^{(n)} - x_+^{(n)} \dfrac{x_i^{(n)}}{x_t^{(n)}} = x_i^{(n)} \dfrac{y_t^{(n)}}{x_t^{(n)}}, & \tilde{y}_j^{(n)} = y_j^{(n)}, \text{ if } x_+^{(n)} < 0, \\[1em] \tilde{x}_i^{(n)} = x_i^{(n)}, & \tilde{y}_j^{(n)} = y_j^{(n)} + x_+^{(n)} \dfrac{y_j^{(n)}}{y_t^{(n)}} = y_j^{(n)} \dfrac{x_t^{(n)}}{y_t^{(n)}}, \text{ if } x_+^{(n)} > 0, \end{cases} \qquad [D.1]$$

where $\tilde{x}_i^{(n)}$ and $\tilde{y}_j^{(n)}$ are the balanced entry and exit counts at stations $i$ and $j$, respectively. The excess entry counts $x_+^{(n)}$ are determined from the total summed entry and exit counts over all stations for the particular observation, according to

$$x_+^{(n)} = x_t^{(n)} - y_t^{(n)}, \qquad [D.2]$$

where

$$x_t^{(n)} = \sum_{i=1}^{S} x_i^{(n)}, \qquad y_t^{(n)} = \sum_{j=1}^{S} y_j^{(n)}.$$

**Appendix E      Test network A**

In the instantaneous-balance model, the travel time between origins and destinations in the network is assumed to be zero. In addition, the in- and outflows are assumed to be balanced for every observation that has been recorded. We create test network A in order to verify that the instantaneous-balance model is generally able to infer the true OD-coefficients from only the entry and exit count observations. The test network consists of $S$ station nodes and observations are assumed to be recorded daily. The underlying link structure is unknown. However, the following assumptions hold: (i) The network is fully connected and all in- and outflow at every station node can be fully observed; (ii) Travel time between nodes is ignored, such that any changes in inflow instantaneously affect changes in outflow; (iii) The total number of entering passengers is equal to the total number of exiting passengers (i.e., flows are balanced) during every observation window.

**E.1 OD-matrix and count data generative process.** The generative process creates the test data from an assumed known (i.e., hypothetical) OD-matrix. The non-negative OD-coefficients are generated for each station $i$ from a symmetric Dirichlet distribution with concentration parameter $c_i$, uniformly sampled between 0 and 2, to create an $(S-1)$-simplex and enforce the sum-to-one constraint,

$$c_i \sim U(0, 2),$$

$$\underset{1 \times (S-1)}{\widetilde{\boldsymbol{\alpha}}_{i\bullet}} \sim \text{Dir}(c_i \mathbf{1}), \quad [\text{E.1}]$$

where $\widetilde{\boldsymbol{\alpha}}_{i\bullet}$ is a row vector of size $(S-1)$. By letting the concentration parameter range up to 2, a stronger degree of similarity between coefficients associated with the same origin is possible, that is, for $c_i > 1$ coefficients concentrate more densely towards the centre of the simplex. The sampled OD-coefficients are mapped to the $i$th row of the OD-matrix in Eq. (1) according to

$$\alpha_{ij} = \begin{cases} \widetilde{\alpha}_{i,j}, & \text{if } i > j, \\ 0, & \text{if } i = j, \\ \widetilde{\alpha}_{i,j-1}, & \text{if } i < j. \end{cases} \quad [\text{E.2}]$$

At every station node $i$ and for every observation $n$, cumulative passenger entry counts $x_i^{(n)}$ are generated from a Negative Binomial distribution with mean parameter $\mu_{x,i}$ and dispersion parameter $\phi \in (0, \infty)$,

$$x_i^{(n)} \sim \text{NegBin}(\mu_{x,i}, \phi) \quad [\text{E.3}]$$

The Negative Binomial distribution is parameterised in the following form

$$P(Z = z | \mu, \phi) = \frac{\Gamma(\phi + z)}{\Gamma(z+1)\Gamma(\phi)} \left(\frac{\phi}{\phi + \mu}\right)^\phi \left(\frac{\mu}{\phi + \mu}\right)^z, \quad z = 0, 1, \dots, \quad [\text{E.4}]$$

where $\Gamma$ is the gamma function. In this way, the variance of the Negative Binomial distribution is defined by

$$\sigma^2 = \text{Var}[z|\mu, \phi] = \mu + \frac{\mu^2}{\phi}. \quad [\text{E.5}]$$

As the dispersion parameter $\phi$ becomes smaller, the variance increases and the distribution becomes more overdispersed; as $\phi$ increases, the more $\sigma^2 \to \mu$ and hence the more the distribution becomes equidispersed.

The generative process views the OD-split coefficients as conditional probabilities, where every split coefficient is the probability of choosing destination $j$ given that a passenger accesses the network at origin $i$. Given the choice probabilities, the generative process creates the cumulative exit count observations in the following way: For every passenger count out of the total entry counts $x_i^{(n)}$ at origin $i$ during observation interval $n$, choose a destination $j$ according to a categorical distribution with probabilities given by $\widetilde{\boldsymbol{\alpha}}_{i\bullet}$, such that for origin $i$ and destination $j$,

$$P(j|\widetilde{\boldsymbol{\alpha}}_{i\bullet}) = \alpha_{ij}, \quad \text{with } j \in \{1, 2, \dots, S\}, \quad [\text{E.6}]$$

where $\widetilde{\boldsymbol{\alpha}}_{i\bullet}$ is the choice probability vector given by the $i$th row of the OD-matrix in Eq. (E.1); Next, the total number of exit counts at the chosen destination are incremented by 1; This process is repeated $x_i^{(n)}$ times for every origin $i$ and observation $n$. The resulting total exit counts at destination $j$ during observation $n$ are denoted as $y_j^{(n)}$.

**E.2 Test data generation.** Test network A consists of $S = 15$ stations. The mean entry count levels across all observation windows and for each station are given in Table E.1. The dispersion parameter of the Negative Binomial model in Eq. (E.3) is initially fixed to $\phi = 10$. Exit counts are subsequently determined according to the aforementioned generative process, and thus by sequentially applying the categorical choice model in Eq. (E.6) for every observation instance and entry count at each station. The number of observation windows is initially set to $N = 30$.

Table E.1. Mean entry count levels for the 15-station instantaneous-balance test network A.

| $i$ | 1 | 2 | 3 | 4 | 5 | 6 | 7 | 8 |
|---|---|---|---|---|---|---|---|---|
| $\mu_{x,i}$ | 600 | 900 | 1400 | 400 | 550 | 650 | 900 | 1000 |
| $i$ | 9 | 10 | 11 | 12 | 13 | 14 | 15 | |
| $\mu_{x,i}$ | 750 | 450 | 200 | 650 | 750 | 1000 | 1300 | |

## Appendix F     Test network B

The average-delay model lifts the assumptions of the instantaneous-balance model, and instead assumes an average travel time between every OD-pair. Moreover, it allows for an in- and outflow imbalance when considering the total entry and exit counts over a truncated observation window. In order to verify that the average-delay model is able to recover the true OD-matrix and in order to evaluate how it compares to the instantaneous-balance model, we devised another test network.

Test network B also consists of $S$ stations. Passengers are able to both enter into and exit from the system at all stations. This time, however, the resolution of the passenger count measurements is increased. Observations are again recorded daily. In addition, every daily observation window is segmented into time bins. In general, the width of the observation window and the number and width of time bins can be tailored to the specific analysis.

The generative process for test network B consists of creating an OD-coefficient matrix, generating time-variant entry count series, and generating the resulting exit count times series according to a multiple-route path choice model. These steps are explained in the upcoming sections, starting with the generation of the OD-coefficient matrix in section F.1, whereas sections F.2 and F.3 delve into the entry count and route choice model.

**F.1 OD-matrix generative process**.

The unknown true OD-matrix is assumed constant across all time bins and observations. It is generated by sampling from a Dirichlet distribution for every origin $i$. The Dirichlet distribution takes a vector $\boldsymbol{c}_i$ of size $(S-1)$, such that

$$\underset{1\times(S-1)}{\widetilde{\boldsymbol{\alpha}}_{i\bullet}} \sim \text{Dir}(\boldsymbol{c}_i) \,. \quad [\text{F.1}]$$

The vector $\boldsymbol{c}_i$ consists of the concentration parameters $c_j$, with $j \in (1, 2, \ldots, S)$, where $j \neq i$, such that $\boldsymbol{c}_i = (c_1, c_2, \ldots, c_{i-1}, c_{i+1}, \ldots, c_{S-1})$. The concentration parameters are sampled once from a uniform distribution, according to

$$c_j \sim U(0, 1) \,. \quad [\text{F.2}]$$

Here, we constrain the concentration parameters between 0 and 1. Since Eq. (F.1) defines a generic Dirichlet distribution, the sparsity of the coefficient samples is controlled by the relative weight of each of the concentration parameters with respect to all other concentration parameters (the magnitude of the concentration parameters only controls the variance of the Dirichlet distribution). In addition, we point out that the value for $c_j$ is only sampled once for every station $j$; That is, for every sample $\widetilde{\boldsymbol{\alpha}}_{i\bullet}$, the concentration parameter vector $\boldsymbol{c}_i$ consists of the same source values $c_j \in \{c_1, c_2, \ldots, c_{S-1}\}$ with $(j \neq i)$. Each concentration parameter represents the attractivity of a specific destination $j$. This is to replicate that certain stations in a network will attract more trips than others; For instance, during the morning commute hours when most passengers will travel from residential areas to business and commercial districts. The vector $\widetilde{\boldsymbol{\alpha}}_{i\bullet}$ is mapped to the $i$th row of the OD-matrix in Eq. (1) according to Eq. (E.2).

**F.2 The entry count hybrid ARIMA-INGARCH model.** Passenger entry counts are assumed as a superposition of a time-dependent trend and a stochastic process distortion. Entry counts are therefore simulated from an auto-regressive integrated moving average (ARIMA) model paired with an integer-valued generalised autoregressive conditional heteroscedasticity (INGARCH) model (Ferland et al., 2006), where we use the *forecast* R-library to simulate from the ARIMA model and the *tscount* R-library (Liboschik et al., 2017) to simulate from the INGARCH model.

The integer-valued generalised autoregressive conditional heteroscedasticity (INGARCH) model (Ferland et al., 2006) describes a count time series $\{Y_t : t \in \mathbb{N}\}$ by modelling its conditional mean as $\mathbb{E}[Y_t|\mathcal{F}_{t-1}] = \lambda_t$, where $\lambda_t$ is a process with $\{\lambda_t : t \in \mathbb{N}\}$. The history $\mathcal{F}_{t-1}$ is the history of the joint process $\{Y_t, \lambda_t, \boldsymbol{Z}_{t+1} : t \in \mathbb{N}\}$ up to time $t$ for $Y_t$ and $\lambda_t$, and $t+1$ for the time-varying r-dimensional covariate vector $\boldsymbol{Z}_t = (Z_{t,1}, \ldots, Z_{t,r})^T$. The general form of the joint process is

$$g(\lambda_t) = \nu_0 + \sum_{k=1}^{p} \nu_k \tilde{g}(Y_{t-i_k}) + \sum_{l=1}^{q} \xi_l g(\lambda_{t-j_l}) + \boldsymbol{\gamma}^T \boldsymbol{Z}_t \,, \quad [\text{F.3}]$$

where $g : \mathbb{R}^+ \to \mathbb{R}$ is a link function and $\tilde{g} : \mathbb{N}_0 \to \mathbb{R}$ is a transformation function. The parameter $\beta_0$ is the intercept. The second right-hand side term in Eq. (F.3) is a regression on past observations $Y_{t-i_1}, Y_{t-i_2}, \ldots, Y_{t-i_p}$ with integers $0 < i_1 < i_2 < \cdots < i_p < \infty$ and $p \in \mathbb{N}_0$. The third term defines a regression on the lagged conditional means $\lambda_{t-j_1}, \lambda_{t-j_2}, \ldots, \lambda_{t-j_q}$, with $0 < j_1 < j_2 < \cdots < j_p < \infty$ and $q \in \mathbb{N}_0$. The parameters $\nu_k$ and $\xi_l$ are the respective weights on regressed lags. The parameters $\boldsymbol{\gamma} = (\gamma_1, \ldots, \gamma_r)^T$ weigh the effects of the covariates.

For the purpose of simulating the entry counts at stations we assume that $g$ and $\tilde{g}$ are equal to the identity, replace the observations $Y_t$ with our notation for the entry counts $x_t^{(n)}$ (the entry counts at time $t$ during observation $n$; for the sake of clarity we omit the station sub-index $i$; However, entry count time series are simulated for every station separately). Moreover, we assume a Negative Binomial (c.f., Eq. (E.4)) entry count likelihood model,

$$x_t^{(n)} | \mathcal{F}_{t-1} \sim \text{NegBin}\big(\lambda_t^{(n)}, \phi\big), \qquad [\text{F.4}]$$

where $\phi \in (0, \infty)$ is the dispersion parameter, such that $\text{Var}[x_t^{(n)} | \mathcal{F}_{t-1}] = \lambda_t + \lambda_t^2 / \phi$. In addition, the covariates are a compound of a trend and a so-called intervention. Interventions model idiosyncrasies in the count time series such as sudden spikes or shifts. In our case, the intervention is solely used to simulate the initial value of the time series. The resulting model for the mean function of the count time series process is

$$\lambda_t^{(n)} = \nu_0 + \sum_{k=1}^{p} \nu_k x_{t-k}^{(n)} + \sum_{l=1}^{q} \xi_l \lambda_{t-l}^{(n)} + \omega_0 \delta_0^{t-\tau} \mathbb{1}(t = \tau) + \boldsymbol{\gamma}^T \boldsymbol{Z}_t, \qquad [\text{F.5}]$$

where $\delta_0^{t-\tau} \mathbb{1}(t = \tau)$ denotes the deterministic intervention covariate, with $\tau$ being the time occurrence and $\delta_0^{t-\tau}$ being the constant decay rate starting from time $\tau$. Since the intervention models the times series' initial value, we fix $\tau = 0$, $\delta_0^{t-\tau} = 0$ (corresponds to a spike), and

$$\omega_0 = \mu_f (1 - \eta). \qquad [\text{F.6}]$$

The parameter $\mu_f$ is the mean of a trend process $F_t$ and $\eta \in (0,1)$ is a scale parameter that captures the effect strength of the trend on the time series. The covariate weights are given by $\boldsymbol{\gamma} = (\omega_0, 1)^T$, with covariate vector $\boldsymbol{Z}_t = (\delta_0^t \mathbb{1}(t = 0), F_t)^T$.

The remaining component to be defined is the trend process. We model the trend according to a seasonal auto-regressive integrated moving average (ARIMA) process. The ARIMA model will be denoted with the common notation as $\text{ARIMA}(p, d, q)(P, D, Q)_m$, where the lower- and uppercase "p, d, q" parameters are non-negative integers defining the order of the autoregressive $p$, differencing $d$, and moving average $q$ components of the non-seasonal part (lowercase) and seasonal part (uppercase), respectively. The span of the periodic season is given by $m$.

Since ARIMA models are defined over all reals and we require that the trend is non-negative, we subtract the simulated trend process' minimum. We denote the resulting process as $H_t$. Next, we introduce the scale parameter $\eta$, such that the trend $F_t$ becomes

$$F_t = \eta H_t + (1 - \eta) \mathbb{E}[H_t], \qquad [\text{F.7}]$$

where $\mathbb{E}[H_t] = \overline{H}_t$ is the mean of the non-negative series $H_t$. In this way, the mean of the trend $F_t$ will be independent of the scale parameter $\eta$. To ensure non-negativity, $\eta$ is constrained between 0 and 1.

**F.3 The multinomial logit path choice model.** Passengers choose between different path alternatives when travelling between an origin and a destination. The travel time to reach their destination varies and thus the time bin when the corresponding exit counts are incremented varies, depending on the selected path alternative. The path choice model is based on a network with a known link structure. Moreover, schedule information is available to generate multiple path alternatives for every OD-pair. Path choice is based on a multinomial logit model and considers the utility (i.e., generalised travel cost) of each path alternatives and generates the probabilities of choosing each alternative in the choice set.

The multinomial logit model (McFadden, 1973) is based on random utility choice models, and aims to determine the choice probabilities within a set of alternatives. Here, the alternatives are the path alternatives between the origins and destinations of the transit network. The utility of each path alternative is equal to the weighted sum of attributes $z_k$ with fixed attribute coefficients $\boldsymbol{\zeta}$. The attributes $z_k = (z_{k,1}, z_{k,2}, z_{k,3}, z_{k,4})$ of path alternative $k$ are the total in-vehicle time $z_{k,1}$, total wait time $z_{k,2}$, total transfer time $z_{k,3}$, and number of transfers $z_{k,4}$. The probability of choosing a particular path $k$ with utility $V_k$ from a set of $K$ possible path alternatives is given by[4]

$$p_k = \frac{e^{V_k}}{\sum_{i=1}^{K} e^{V_i}}, \quad [\text{F.8}]$$

with

$$V_i = \boldsymbol{\zeta}^T \boldsymbol{z}_i. \quad [\text{F.9}]$$

The choice between the $K$ alternatives is generated from a categorical choice model, according to

$$P(k|\boldsymbol{p}) = p_k, \quad \text{with } k \in \{1, 2, \dots, K\}, \boldsymbol{p} = (p_1, p_2, \dots, p_K). \quad [\text{F.10}]$$

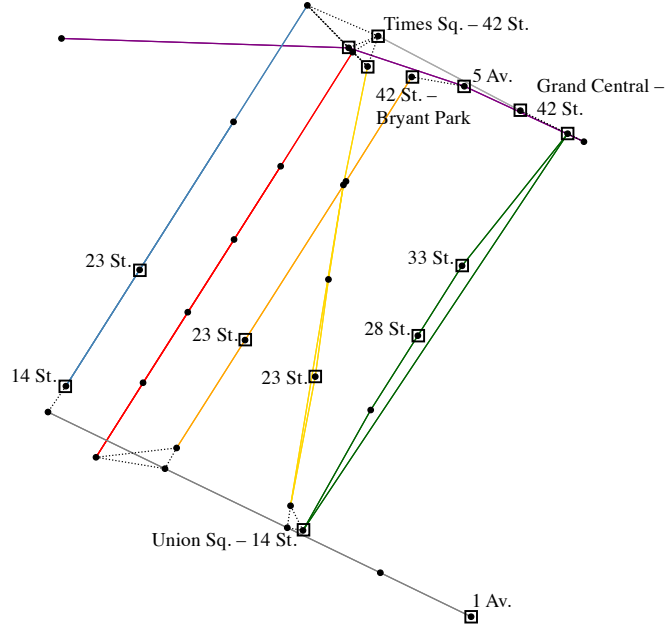

**Figure F.1.** Test network B – A subcomponent of the NYC subway system; Colors indicate the different subway lines. Dotted black lines indicate transfer footpaths. The annotated stations are assumed to be the only access and egress locations of the network.

**F.4 Test data generation.** The average-delay test case is based on a subcomponent of the NYC subway network. The chosen subcomponent is shown in Fig. F.1. The subcomponent network consists of 35 stations, 8 lines, and 39 transfer linkways (i.e., footpaths). Out of the 35 stations, we pick $S = 15$ stations for which we generate $N$ entry count time series, following the generative process described earlier. The departure window starts at $t_0 = 7$ AM, ends at 9 AM, and is discretised into 5-minute time bins (i.e., $T = 24$).

After generating the OD-matrix according to Eq. (F.1), any coefficients corresponding to station pairs connected by a direct transfer path are set to zero. The corresponding OD-matrix row is re-normalized to ensure the simplex constraint is satisfied.

Entry count series are simulated for each of the 15 stations over a fixed-length window and at pre-defined time intervals, according to the hybrid ARIMA-INGARCH process in section F.2. The trend covariate $F_t$ (c.f., Eq. (F.7)) is generated once for each station; Next, we generate $N$ count series for each station, according to the INGARCH model in Eq. (F.5). The intercept $\nu_0$ is fixed at 1. The observation and mean regression parameter values are $\nu_1 = 0.1$ and $\nu_2 = 0.05$, and $\xi_1 = 0.1$ and $\xi_2 = 0.05$, respectively. The trend is simulated from an $\text{ARIMA}(1,1,1)(1,1,1)_T$ model, where $T$ is the number of simulated time bins and corresponds to the number of time bins contained in the departure window defined in Section 2.3. The coefficients of the autoregressive and moving average components are set to 0.8 and 0.9 for the non-seasonal part, and 0.5 and -0.5 for the seasonal part. The adjustable parameters are the trend scale $\eta$ and the count dispersion $\phi$. Fig. F.2 shows examples of the entry count time series mean and percentile range for different $\eta$ and $\phi$ parameter values. In this example, entry count series are simulated $N = 100$ times over an assumed 2-hour window, segmented into 5-minute intervals, resulting in $T = 24$ time bins. We simulate $T + 1$ time bins and truncate the first sample to avoid a fixed starting value.

---

[4] In order to avoid numerical overflow due to the exponential terms in Eq. (F.8), we use an arithmetic reformulation using the log-sum of exponential terms. By first determining the maximum utility $V_{\max}$, the expression in Eq. (F.8) can be rewritten to $p_k = \exp[V_k - (V_{\max} + \log(\sum \exp(V_i - V_{\max})))]$.

Path alternatives are generated based on the schedule information published by the Metropolitan Transport Authority (MTA) of NYC (Metropolitan Transport Authority, 2019b), whereby all path alternatives that fall within a 10-minute margin from the earliest arrival path are considered. The schedule information is available in the General Transit Feed Specification (GTFS) format and parsed into queryable data structures. We fix the proxy departure time to 8 AM – i.e., any passenger entering the system during the 7 AM to 9 AM window will see the same path alternatives corresponding to a departure time of 8 AM. The path alternatives provide all necessary route attribute information required for the multinomial logit model in Eq. (F.8), with attribute coefficients fixed to $\zeta = (-0.002, -0.006, -0.002, -1.0)^T$.

The generative process simulation proceeds to increment the exit counts according to the entry counts at each origin, the departure time bins, given destination, and path choice of every passenger entering the system. The arrival window starts at $t_1 = 8$ AM. The maximum travel time through the test network is about 40 minutes. The start time of the departure time window at $t_0 = 7$ AM thus provides sufficient delay to capture passengers who are arriving at their destination station at exactly 8 AM and later.

Once $N$ entry count time series at each of the $S$ stations are simulated, the generative model iterates over every station, observation, time bin, and entry count. For every entry count increment, a passenger chooses a destination according to the categorical choice model in Eq. (E.6), where the choice probabilities are given by the OD-matrix coefficients generated from Eq. (F.1). Next, the passenger chooses a path alternative according to the choice model in Eq. (F.10). The departure time bin, chosen path, and corresponding travel time decide when the traveler arrives at their destination. If the arrival time bin is within the arrival window, the exit counts during that respective arrival time bin are incremented. This process repeats until all entry counts are processed.

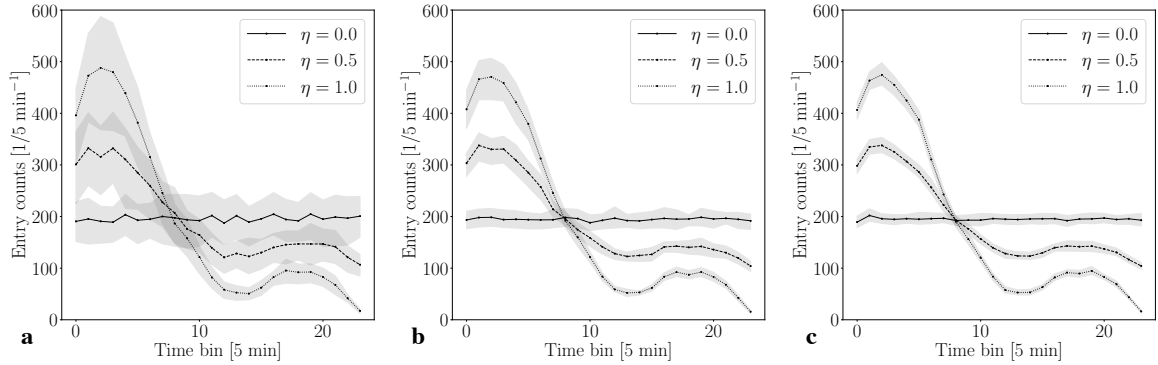

**Figure F.2.** The mean and 25th to 75th percentile range of the entry count time series at a single station for 24 time bins and evaluated over 100 observations, based on the hybrid ARIMA-INGARCH generative entry count model. Dots connected by solid lines indicate the mean, and grey-shaded areas indicate the percentile range. The graphs correspond to different values of the scale parameter $\eta$ and Negative Binomial dispersion parameter $\phi$; (a) $\phi = 10$; (b) $\phi = 100$; (c) $\phi = 1000$.

## Appendix G     Raw data processing and filtering

The turnstile count records of the NYC subway system contain entry and exit counts in 4-hour intervals and erroneous measurements. Therefore, the raw count records of the number of passengers accessing and egressing at stations must be cleaned and processed before they can be used as input data in the OD-inference model. First, they are up-sampled and aggregated over the desired observation windows. Despite this pre-processing of the data, the count records are still distorted by noise, missing data, and outliers. Thus, in a second step, the data are filtered to impute new estimates for missing and corrupted data points.

**G.1 Count up-sampling.** The first step in treating the data records is to up-sample the entry and exit counts to the desired recording intervals. A critical property is to up-sample the counts conservatively in order to preserve the total number of counts over the original 4-hour intervals. We do this by monotonically interpolating the cumulative counts at points of the desired sampling frequency, and differencing the interpolated values to obtain the up-sampled counts. This way, when integrating over the up-sampled counts, the original cumulative counts can be recovered. The interpolated values are, however, not integer count values, but positive reals. The interpolator of the cumulative counts thus needs to be monotonically increasing. Additionally, the interpolator needs to be $C^2$-continuous to ensure smoothness of the differenced function values. Rational Quadratic Spline interpolation (Delbourgo and Gregory, 1983) is such an interpolator. We implement the interpolation routines into a custom up-sampling method, that finds contiguous segments of count records, monotonically interpolates their cumulative counts, and up-samples the segments to the desired frequency. Here, we up-sample the counts to 5-minute intervals and aggregate them over the respective observation windows before applying the instantaneous-balance inference model – Up-sampling to 1-hour intervals would give the same resulting counts; the 5-minute interval count data, however, allow for a more flexible use in testing the average-delay inference model.

**G.2 Filtering and missing data imputation.** Once the up-sampled data are aggregated over the desired observation window, they are still interspersed with missing and corrupted data, requiring a filtering procedure. The filtered data are obtained from a de-noising autoencoder neural network implemented using the *tensorflow* package for Python. The unsupervised learning algorithm aims to minimise the target loss, defined as the sum of the reconstruction loss and a regularisation loss. The reconstruction loss measures the mean squared error between the (normalised) count observations and the neural network predictions. The regularisation loss is defined by an $L_2$-regularisation term over the hidden layers' weights to prevent overfitting. The regularisation parameter is set to 0.0001.

The input layer includes observation neurons that feed in count observations, as well as masking neurons that feed in a Boolean mask to identify missing data and dynamically-changing outliers. The number of observation and masking neurons depends on the number of stations and a specified sub-frame width. The neural network has three hidden layers. The input layer is connected to the first hidden layer via a drop-out layer with a 30% de-noising drop-out rate. The first hidden layer uses a rectified linear activation function. The output layer has the same number of neurons as the input layer. We choose 150 neurons for the first and third hidden layer, and 75 neurons for the second hidden layer.

Training the autoencoder neural network requires a defined number of epochs. Every epoch consists of multiple iterations to train the neural network. Every iteration draws a new batch of multiple training samples of contiguous records with specified sub-frame width $M$. The sub-frame width defines the number of contiguous records in every sample. The samples are drawn at random from the full set of recorded data. The batches of training samples are fed into the neural network and are processed forward and backward to train the neural network while minimising the loss due to reconstruction error and weighing in the regularisation error. Every processed batch completes one iteration.

Once all iterations have been completed, the neural network is tested by evaluating its predictions versus the observed count records. Prior to testing, the predictions are filtered through a moving average window from the first to the last record. The width of the moving average window is chosen to be the same as the sub-frame width $M$. These filtered predictions are then used to determine the squared error for non-missing data. We use the squared error to determine the upper percentile of observations that qualify to be outliers, based on an estimated minimum fraction of outliers in the data. The estimated fraction of outliers is set to 5%. To correct for potentially having misidentified outliers, an outlier drop-out rate is fixed at 0.1 to randomly re-qualify data points as valid. According to the identified outliers, the outliers mask is changed to reflect whether a data point is flagged as an outlier or not. Next, outliers are corrected by a weighted average of the observation value and the prediction. The weights are set to 0.5, effectively determining the average between the observation and prediction. In addition, missing data points are imputed with a weighted average of the previous epoch's and current epoch's prediction.

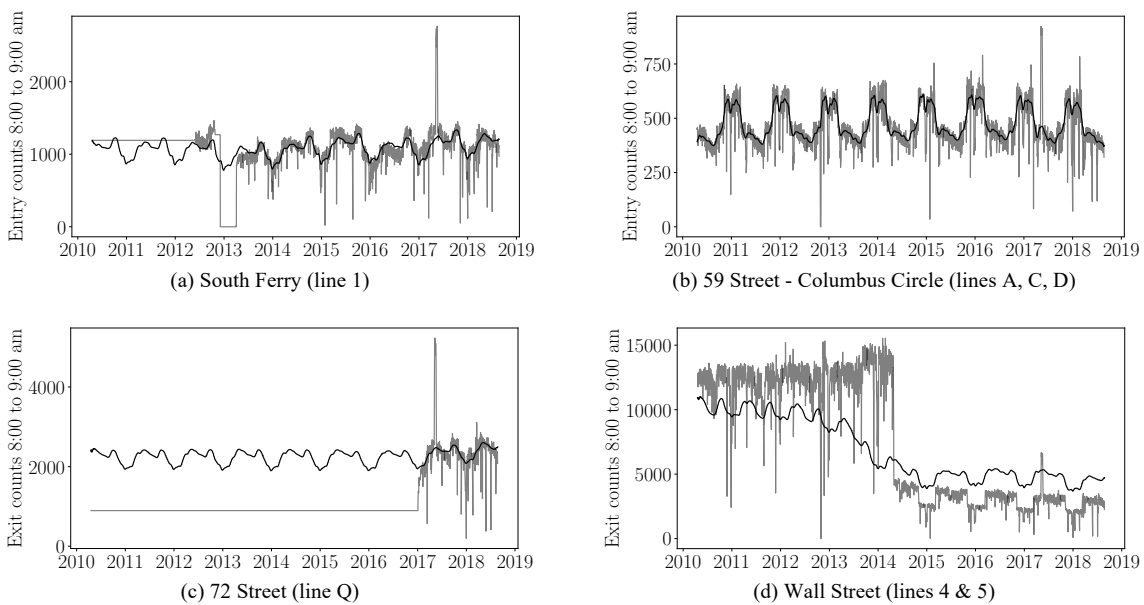

**Figure G.1.** Filtered entry and exit counts for different stations in the NYC subway network with varying degree of missing data and outliers (missing data are generated as detailed in Appendix G). Grey curves plot the measured counts, black curves plot the filtered counts.

The weight is set to 0.5. Once these steps are complete, the algorithm steps into the next epoch and repeats – this time with the corrected values for the outliers and missing data.

The batch size $B$ is specified according to user input. The total number of records is denoted as $N_{nn}$. The number of iterations per epoch is defined by the sub-frame width and batch size, according to $(N_{nn} - M)/B$, rounded to the nearest integer. Thus, during every epoch, the neural network is trained over $(N_{nn} - M)/B$ iterations. Each iteration draws one batch consisting of $B$ samples of count records. Every sample consists of $M$ contiguous records. In this case, we choose $M = 50$ records and a batch size of $B = 15$ samples for every batch. The number of input neurons is equal to $2S \times M$, where S is the number of stations. The number of epochs is set to 100.

Fig. G.1 shows examples of the processed time series of the number of passengers who entered and exited at selected stations within the 8:00-to-9:00 AM window from 2010 to 2018. The grey curves are the pre-processed time series. The black curves are the filtered time series, determined from the autoencoder neural network.

**Appendix H    Effect of model regularisation**

The instantaneous-balance Bayesian inference model and QP optimisation model are run on test samples with either 5, 10, 30, or 100 observations derived according the generative process for test network A in Appendix E. Next, the inferred OD-coefficient estimates are used to predict the exit counts for out-of-sample validation data according to the uncorrected instantaneous-balance model formulation $Y = XA$. The validation data consisting of 10000 out-of-sample observations are denoted as $y$ and derived from the same generative process as the sampled observations for test network A. However, they are not used in the estimation of the OD-coefficients. The Bayesian model predictions $\hat{\mu}_y$ are determined by inserting the OD-coefficient posterior means into $A$; The QP model predictions $\hat{y}$ are determined by inserting the OD-coefficient point estimates. This procedure is carried out twice; Once, by determining OD-coefficient estimates according to the regularized formulations in Eq. (3) for the Bayesian inference model and Eq. (A.1) for the QP inference model; And again, by using the otherwise same formulations except that the regularization terms are omitted. Fig. H.1 shows that the MSE of the MCMC and of the QP estimate decreases as the number of observations used for the OD-estimation increases, given that more observations provide more evidence of the underlying, latent OD-matrix. Furthermore, Fig. H.1 confirms that the regularised solution outperforms the non-regularised solution when tested against validation data.

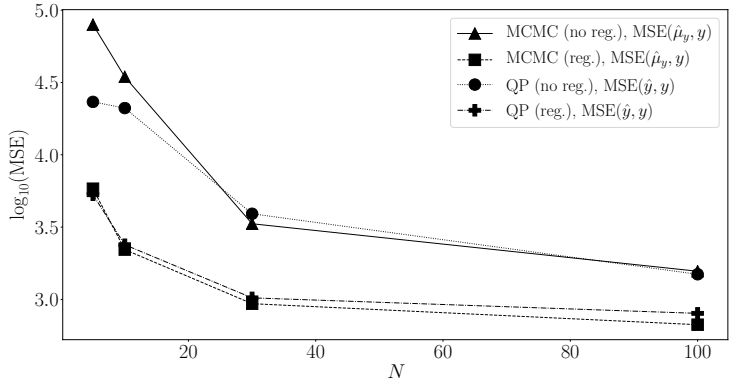

**Figure H.1.** The influence of model regularisation and the number of observations $N$ on the instantaneous-balance model predictions, measured in terms of the mean squared error (MSE) between the MCMC-based (and QP-based) predictions and out-of-sample validation data for test network A. The MSE is determined across all 15 stations and 10000 validation data points.

**Appendix I    Sensitivity analysis**

We assess the correlation between accuracy and precision of the instantaneous-balance and average-delay estimates, and their sensitivity to the observation window width $w$, the trend scale $\eta$ and the count dispersion $\phi$ using a metamodel/emulator, which describes the relation between each statistic (i.e., the response variable) and these parameters (i.e., the regressors) according to the linear model

$$q = \beta_0 + \beta_1 w + \beta_2 \eta + \beta_3 \phi + \epsilon, \quad [\text{I.1}]$$

where $q$ is a placeholder for either the accuracy or precision statistic (i.e., either $MSE(\mu_{\alpha_{ij}}, \alpha_{ij})$ or $\hat{\mu}_{HPD}$), $w$ denotes the observation window width, and $\eta$ and $\phi$ are the trend scale and Negative Binomial dispersion. We fit the model via the MinMax-normalised regressors and by determining the least-squares estimate of the coefficients $\beta_i$. In this way, we construct four metamodels for the accuracy and precision of the instantaneous-balance and average-delay model. For each metamodel, we test for the statistical significance of its coefficients and determine the first-order and total Sobol' sensitivity indices. The Sobol' indices are based on a Monte Carlo estimate with 10000 samples from a uniform distribution for each parameter and 1000 bootstrap replicates. All corresponding statistics regarding the correlation metrics, metamodel parameters, and Sobol' indices are collected in Table I.1.

Table I.1 confirms that increasing the window width improves significantly both models' accuracy and precision. Indeed, accuracy and precision are negatively correlated with the window width, i.e., the larger the window width, the smaller the estimates' MSE and mean HPD. P-values indicate high significance of the window width regressor and Sobol' indices indicate that the accuracy exhibits large fist-order sensitivities, i.e., large variance of the conditional expectation of the accuracy with respect to the window width and relative to the total variance, and little indication of any higher-order interactions, i.e., total indices are similar to first-order indices. The underlying assumption for both the generative and inferential model is that the OD-matrix stays constant throughout the observation window. Consequently, increasing the window width will increase the number of time bins contained within the observation window and, as a result, will add more data points and thus evidence for the average-delay model to infer the OD-coefficients. Moreover, as the observation window becomes wider, a larger proportion of passenger trips occur within the window itself, which complies with the instantaneous-balance assumption and, thus, improves the instantaneous-balance model's accuracy and precision.

Table I.1 also confirms that the average-delay model does not exhibit significant sensitivity to the trend scale. Because the average-delay model explicitly includes the travel time between stations, any behaviour in the trend of the entry counts at one station is mapped to the exit counts by a lagged time bin at another station. Since the trend stays unchanged for all observations, this mapping will be repetitive and not contribute any new information to improve the inference, whether or not the trend strongly persists. In contrast, the instantaneous-balance model is influenced by the strength of the time series trend, and both the accuracy and precision deteriorate as $\eta$ increases. The instantaneous-balance model ignores the travel time between stations and expectedly will perform best if steady conditions persevere over the observation window. However, the stronger the trend, the less the steady-state balance of the entry flow and exit flow is maintained. As a result, time aggregation over the limited window will induce larger errors to the instantaneous-balance assumption, which consequently degrades the accuracy and precision of the instantaneous-balance model.

Finally, the accuracy and precision of the OD-coefficient estimates improve as the count observations become more dispersed. This is in line with the analysis of the instantaneous-balance model in Section 4.1. Indeed, both the average-delay and the instantaneous-balance model rely on the inherent variability of the entry and exit count observations. The more pronounced the exit-entry count patterns change, the stronger they correlate between the exit and entry count observations, the stronger the evidence is of a possible linkage between particular OD-pairs. Under the assumption that the latent OD-matrix does not change, a stronger dispersion in the entry counts will propagate into a stronger dispersion of the exit counts. As a result, inferring the associated OD-coefficients will become easier.

**Table I.1.** The sensitivity of the models' estimation accuracy and precision to test parameters. The table lists the correlations between the test parameter values and the models' estimation accuracy and precision, and the coefficients of the fitted model in Eq. (I.1) along with their p-values as well as the test parameters' first-order and total Sobol' indices ($S$ and $S_T$, respectively). Columns with header I correspond to the instantaneous-balance model; Columns with header II correspond to the average-delay model. Positive correlation with respect to the window width, trend scale, or dispersion parameter means that a decrease in either test parameter will improve the model's accuracy or precision, and vice versa. The variance/dispersion of the samples drawn from the negative binomial distribution is inversely proportional to the dispersion parameter $\phi$.

| | \multicolumn{8}{c}{Accuracy} | | | | | | | |
|---|---|---|---|---|---|---|---|---|
| | Window width, $w$ | | Trend scale, $\eta$ | | Dispersion, $\phi$ | | Intercept, $\beta_0$ | |
| | I | II | I | II | I | II | I | II |
| Corr. | -0.8986 | -0.6378 | 0.2195 | -0.0645 | 0.1733 | 0.3978 | | |
| Coeff. | -0.0050 | -0.0022 | 0.0011 | -0.0002 | 0.0008 | 0.0011 | 0.0076 | 0.0030 |
| p-val. | $4.65 \cdot 10^{-16}$ | $4.68 \cdot 10^{-6}$ | $8.70 \cdot 10^{-4}$ | $5.82 \cdot 10^{-1}$ | $6.70 \cdot 10^{-3}$ | $1.69 \cdot 10^{-3}$ | $< 2 \cdot 10^{-16}$ | $5.98 \cdot 10^{-11}$ |
| $S$ | 0.9273 | 0.7832 | 0.0450 | 0.0105 | 0.0261 | 0.2164 | | |
| $S_T$ | 0.9235 | 0.7748 | 0.0471 | 0.0067 | 0.0245 | 0.2097 | | |
| | \multicolumn{8}{c}{Precision} | | | | | | | |
| | Window width, $w$ | | Trend scale, $\eta$ | | Dispersion, $\phi$ | | Intercept, $\beta_0$ | |
| | I | II | I | II | I | II | I | II |
| Corr. | -0.4884 | -0.7237 | 0.1787 | 0.0200 | 0.7179 | 0.5347 | | |
| Coeff. | -0.0273 | -0.0716 | 0.0092 | 0.0018 | 0.0336 | 0.0442 | 0.1281 | 0.0934 |
| p-val. | $1.20 \cdot 10^{-6}$ | $1.02 \cdot 10^{-10}$ | $3.60 \cdot 10^{-2}$ | $7.97 \cdot 10^{-1}$ | $5.00 \cdot 10^{-10}$ | $7.32 \cdot 10^{-8}$ | $< 2 \cdot 10^{-16}$ | $< 2 \cdot 10^{-16}$ |
| $S$ | 0.3772 | 0.7267 | 0.0301 | 0.0029 | 0.5823 | 0.2814 | | |
| $S_T$ | 0.3730 | 0.7187 | 0.0427 | 0.0005 | 0.5754 | 0.2729 | | |

## Appendix J  Critical parameter combinations

The best- and worst-cases reported in Fig. 7, for $N = 100$ number of observations, identify the critical combinations of test parameters that produce the smallest and largest estimation errors and uncertainties. Both the instantaneous-balance and average-delay model exhibit the largest error and uncertainty for an observation window with $w = 5$ minutes, trend scale $\eta = 1$, and dispersion $\phi = 1000$. For this worst-case condition, both models equally struggle to identify the true OD-coefficients. The corresponding estimates are visualized in Fig. J.1a. In contrast, the instantaneous-balance model returns the best estimates for $w = 1$ hour, $\eta = 0$, and $\phi = 10$, shown in Fig. J.1b. Fig. J.1c shows the results subject to the average-delay model's best-case conditions with $w = 1$ hour, $\eta = 1$, and $\phi = 10$. This confirms previous analyses in Appendix I, that establishes that longer observation windows and larger count dispersion improve the accuracy and precision of the estimates. Moreover, it supports the reasoning that the instantaneous-balance model favours flat trends, whereas the average-delay model is weakly impacted by the trend scale.

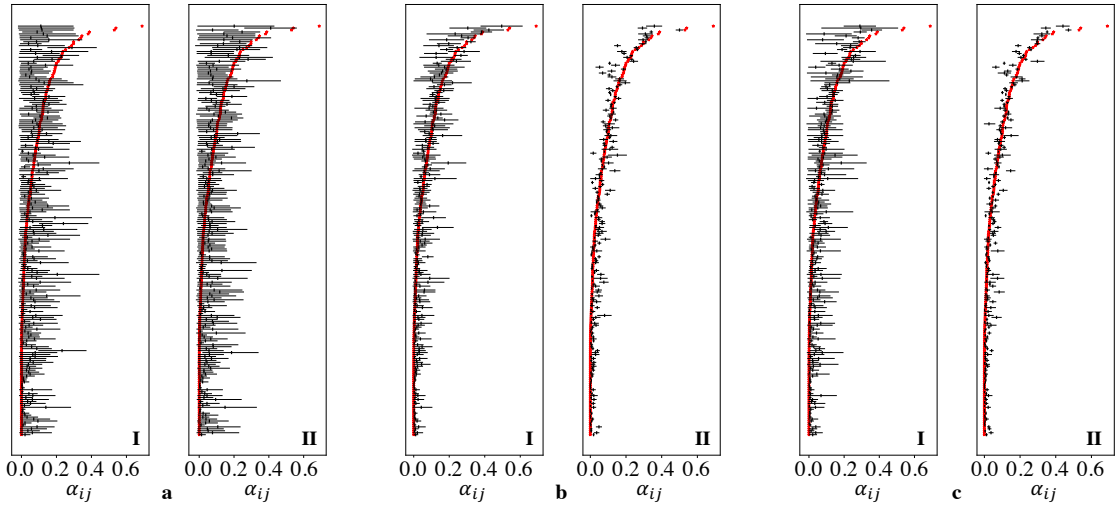

**Figure J.1.** The worst and best-case estimation results for the instantaneous-balance and average-delay model. The red stars are the true OD coefficient values, the vertical and horizontal bars visualise the mean and HPD interval of the estimates. Coefficients are sorted from lowest to greatest along the vertical axis. (a) Both models' worst-case; $w = 5$ min, $\eta = 1$, $\phi = 1000$. (b) The instantaneous-balance model best-case; $w = 1$ hr, $\eta = 0$, $\phi = 10$. (c) The average-delay model best-case; $w = 1$ hr, $\eta = 1$, $\phi = 10$. The left- and right-hand depictions in each sub-figure juxtapose the instantaneous-balance next to the average-delay model results.

## Appendix K  OD-flow in the NYC subway system

Fig. K.1 illustrates the NYC subway OD-matrices consisting of the sample means of the OD-coefficient posteriors mapped onto the geographical locations of the NYC subway network. Fig.'s K.1a to K.1c show that the attractivity of stations changes throughout the day – whereas the largest proportions of passenger trips are bound for the districts in Lower, Midtown, and Upper Manhattan during the morning, the proportions of trips bound for the surrounding boroughs gradually increase by midday, and almost reverse during the evening.

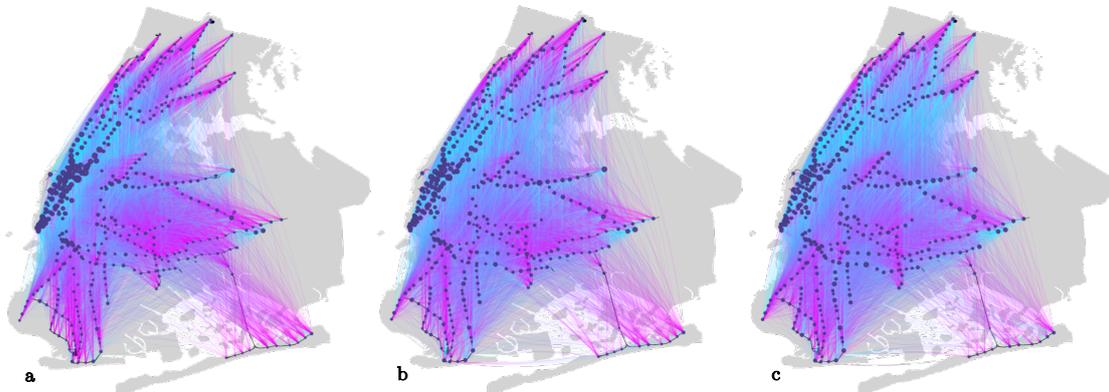

**Figure K.1.** Map representation of the OD-coefficient posterior means in Fig. 9 for the (a) morning peak, (b) midday off-peak, and (c) evening peak window. The canvas map shows the NYC area contour along with the locations of subway tracks and stations. Every OD-coefficient with $\hat{\mu}_{\alpha,ij} > 0.003$ is plotted as an arc between the corresponding station locations. The colour gradient (magenta to blue) indicates the direction (origin: magenta; destination: blue). The OD-coefficient posterior mean defines the width of each arc. The size of the station node markers indicates the total in-degree of the node, i.e., the sum of inbound OD-coefficient means.

Along with the changes in relative OD-demand presented in Section 4.3, the absolute number of passengers travelling between stations gradually changes throughout the day. Fig. K.2 shows an estimate of the absolute passenger demand between all OD-pairs, given the OD-coefficient posterior means and the average weekday passenger access counts between May 2nd and May 13th, 2016. We take the 1-hour time interval defined for each observation window to accumulate the total access counts at each station. The plotted values in Fig. K.2 are computed from $X\bar{A}$, where $X_{N\times S}$ contains the aggregated access counts over the 1-hour interval and $\bar{A}_{S\times S}$ is the OD-matrix consisting of the OD-coefficient posterior means. Fig. K.2 supports the finding of Fig. 10 in Section 4.3, i.e., during the morning peak, a group of only few stations attract a large number of passenger trips from stations all across the network, visible again by the vertical bands in fig. K.2a. By the afternoon off-peak in Fig. K.2b, this dedicated OD-demand gradually reduces and spreads to more stations in the network, that is, the vertical bands become more diffused. During the evening peak in Fig. K.2c, the dedicated OD-demand vanishes and numerous stations throughout the network attract passenger trips from across the network.

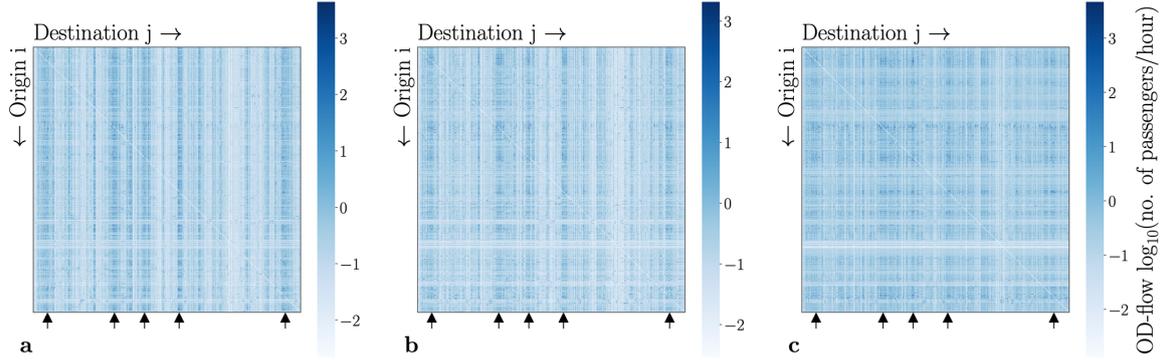

**Figure K.2.** O-D demand matrices corresponding to the (a) morning peak, (b) midday off-peak, and (c) evening peak window. The estimates are based on the sample means of the OD-coefficient posterior and the weekday average total passenger entry count at each station of the NYC subway network during the respective 1-hour time intervals between 2$^{nd}$ May until 13$^{th}$ May 2016. Vertical arrows below each plot mark stations that consistently observe more than 3000 exiting passengers per hour. Rows and columns are sorted according to the geographic locations of stations and the stop patterns of services running through them.

### Appendix L  The effective sample size

As much as Markov chains provide desirable properties to be able to sample from the posterior distribution, their inherent stochastic dependence causes their samples to be autocorrelated. An artefact of this correlation is that the standard errors and uncertainties of parameter estimates increase. The effective sample size $N_{\text{eff}}$ can be used to measure the degree to which the estimates are influenced by the samples' autocorrelation (Geyer, 2011). Here, we report the ratio with respect to the autocorrelated samples $N_{\text{eff}}/N_{MC}$ – The closer this ratio is to 1.0, the more the estimates' uncertainty can be expected to not stem from the samples' inherent autocorrelation. Since the autocorrelation for a chain with multiple parameters and joint probability function cannot be directly calculated (the joint probability function is the posterior density of interest), *Stan* (Carpenter et al., 2017) estimates the autocorrelation and thus effective sample size from a Fast Fourier Transform (FFT) procedure and combines the estimates from multiple chains. In the case of the NYC OD-estimation results, we use a thinned sample of $N_{MC} = 1000$ MCMC draws for the estimation of the effective sample size.